\documentclass[conference]{IEEEtran}
\usepackage[utf8]{inputenc}
\usepackage[english]{babel}
\usepackage[ruled, vlined]{algorithm2e}
\usepackage{algpseudocode}
\usepackage{academicons}
\usepackage{amssymb}
\usepackage{array}
\usepackage{arydshln}
\usepackage{blkarray, bigstrut}
\usepackage{bm}
\usepackage{booktabs}
\usepackage{braket}
\usepackage{cite}
\usepackage{cuted}
\usepackage{enumitem}
\usepackage{graphicx}
\usepackage[mathscr]{euscript}
\usepackage[switch]{lineno}
\usepackage{mathtools}
\usepackage{multirow}
\usepackage{multicol}
\usepackage{mathtools}
\usepackage{xfrac}
\usepackage{framed} 
\usepackage[framed]{ntheorem}
\usepackage{pgfplots}
\usepackage{qtree}
\usepackage{adjustbox}
\usepackage{pifont}
\usepackage{rotating}
\pgfplotsset{compat = 1.12}
\usepackage{stfloats}
\usepackage{subcaption}
\usepackage{url}
\usepackage{tikz}
\usepackage[framemethod=TikZ]{mdframed}
\usetikzlibrary{patterns}
\usetikzlibrary{spy}
\usetikzlibrary{shapes, backgrounds,calc, snakes}
\usetikzlibrary{decorations.pathreplacing}
\usetikzlibrary{decorations.text}
\usetikzlibrary{matrix}
\usepackage{fancybox}
\usepackage{scalerel}
\usetikzlibrary{svg.path}
\usepackage{xcolor}
\usepackage[normalem]{ulem}
\usepgfplotslibrary{fillbetween}
\usepgfplotslibrary{groupplots}
\usetikzlibrary{shapes.geometric,shapes.arrows,decorations.pathmorphing}
\usetikzlibrary{matrix,chains,scopes,positioning,arrows,fit}

\definecolor{orcidlogocol}{HTML}{A6CE39}
\tikzset{
  orcidlogo/.pic={
    \fill[orcidlogocol] svg{M256,128c0,70.7-57.3,128-128,128C57.3,256,0,198.7,0,128C0,57.3,57.3,0,128,0C198.7,0,256,57.3,256,128z};
    \fill[white] svg{M86.3,186.2H70.9V79.1h15.4v48.4V186.2z}
                 svg{M108.9,79.1h41.6c39.6,0,57,28.3,57,53.6c0,27.5-21.5,53.6-56.8,53.6h-41.8V79.1z M124.3,172.4h24.5c34.9,0,42.9-26.5,42.9-39.7c0-21.5-13.7-39.7-43.7-39.7h-23.7V172.4z}
                 svg{M88.7,56.8c0,5.5-4.5,10.1-10.1,10.1c-5.6,0-10.1-4.6-10.1-10.1c0-5.6,4.5-10.1,10.1-10.1C84.2,46.7,88.7,51.3,88.7,56.8z};
  }
}

\newcommand\orcidicon[1]{\href{https://orcid.org/#1}{\mbox{\scalerel*{
\begin{tikzpicture}[yscale=-1,transform shape]
\pic{orcidlogo};
\end{tikzpicture}
}{|}}}}

\newcommand{\argmax}{\arg\!\max}

\definecolor{bblue}{rgb}{0.12392, 0.0490, 0.9588}
\definecolor{sskyblue}{rgb}{0.1529, 0.5882, 0.9216}
\definecolor{ggreen}{rgb}{0.7098, 0.95, 0.40781}
\definecolor{yyellow}{rgb}{0.9765, 0.9804, 0.0784}

\definecolor{ycolor0}{HTML}{313695}
\definecolor{ycolor1}{HTML}{4575b4}
\definecolor{ycolor2}{HTML}{74add1}
\definecolor{ycolor3}{HTML}{abd9e9}
\definecolor{ycolor4}{HTML}{707068}
\definecolor{ycolor5}{HTML}{990000}
\definecolor{ycolor6}{HTML}{ef6548}
\definecolor{ycolor7}{HTML}{fdbb84}

\definecolor{color0}{HTML}{FF0147}
\definecolor{color1}{HTML}{F400DC}
\definecolor{color2}{HTML}{BA0DFF}
\definecolor{color3}{HTML}{5700E8}
\definecolor{color4}{HTML}{0B03FF}
\definecolor{color5}{HTML}{0957F4}
\definecolor{color6}{HTML}{03B3FF}
\definecolor{color7}{HTML}{08E8DA}
\definecolor{color8}{HTML}{07FF8E}
\definecolor{color9}{HTML}{51FF0A}

\definecolor{p1}{rgb}{1, 0.0667, 0}
\definecolor{p2}{rgb}{1, 0.24, 0}
\definecolor{p3}{rgb}{1, 0.349, 0}
\definecolor{p4}{rgb}{1, 0.490, 0}
\definecolor{p5}{rgb}{1, 0.631, 0}
\definecolor{p6}{rgb}{1, 0.792, 0}
\definecolor{p7}{rgb}{1, 0.933, 0}

\makeatletter
\def\newmaketag{%
	\def\maketag@@@##1{\hbox{\m@th\normalfont\normalsize##1}}%
}
\makeatother

\DeclareMathSizes{10pt}{8pt}{6pt}{5pt} 	
\makeatletter
\let\save@mathaccent\mathaccent
\newcommand*\if@single[3]{%
  \setbox0\hbox{${\mathaccent"0362{#1}}^H$}%
  \setbox2\hbox{${\mathaccent"0362{\kern0pt#1}}^H$}%
  \ifdim\ht0=\ht2 #3\else #2\fi
  }
\newcommand*\rel@kern[1]{\kern#1\dimexpr\macc@kerna}
\newcommand*\widebar[1]{\@ifnextchar^{{\wide@bar{#1}{0}}}{\wide@bar{#1}{1}}}
\newcommand*\wide@bar[2]{\if@single{#1}{\wide@bar@{#1}{#2}{1}}{\wide@bar@{#1}{#2}{2}}}
\newcommand*\wide@bar@[3]{%
  \begingroup
  \def\mathaccent##1##2{%
    \let\mathaccent\save@mathaccent
    \if#32 \let\macc@nucleus\first@char \fi
    \setbox\z@\hbox{$\macc@style{\macc@nucleus}_{}$}%
    \setbox\tw@\hbox{$\macc@style{\macc@nucleus}{}_{}$}%
    \dimen@\wd\tw@
    \advance\dimen@-\wd\z@
    \divide\dimen@ 3
    \@tempdima\wd\tw@
    \advance\@tempdima-\scriptspace
    \divide\@tempdima 10
    \advance\dimen@-\@tempdima
    \ifdim\dimen@>\z@ \dimen@0pt\fi
    \rel@kern{0.6}\kern-\dimen@
    \if#31
      \overline{\rel@kern{-0.6}\kern\dimen@\macc@nucleus\rel@kern{0.4}\kern\dimen@}%
      \advance\dimen@0.4\dimexpr\macc@kerna
      \let\final@kern#2%
      \ifdim\dimen@<\z@ \let\final@kern1\fi
      \if\final@kern1 \kern-\dimen@\fi
    \else
      \overline{\rel@kern{-0.6}\kern\dimen@#1}%
    \fi
  }%
  \macc@depth\@ne
  \let\math@bgroup\@empty \let\math@egroup\macc@set@skewchar
  \mathsurround\z@ \frozen@everymath{\mathgroup\macc@group\relax}%
  \macc@set@skewchar\relax
  \let\mathaccentV\macc@nested@a
  \if#31
    \macc@nested@a\relax111{#1}%
  \else
    \def\gobble@till@marker##1\endmarker{}%
    \futurelet\first@char\gobble@till@marker#1\endmarker
    \ifcat\noexpand\first@char A\else
      \def\first@char{}%
    \fi
    \macc@nested@a\relax111{\first@char}%
  \fi
  \endgroup
}

\definecolor{color6}{HTML}{03B3FF}
\pgfplotsset{major grid style={densely dotted,gray!50!gray}}
\definecolor{dcolor1}{HTML}{253494}
\definecolor{dcolor2}{HTML}{636363}

\definecolor{dcolor3}{HTML}{fcbba1}
\definecolor{dcolor4}{HTML}{fb6a4a}
\definecolor{dcolor5}{HTML}{cb181d}
\definecolor{dcolor6}{HTML}{67000d}

\definecolor{dcolor7}{HTML}{9ecae1}
\definecolor{dcolor8}{HTML}{4292c6}
\definecolor{dcolor9}{HTML}{08519c}

\begin{document}
\title{\LARGE{Sequential Parametric Optimization for Rate-Splitting Precoding in Non-Orthogonal Unicast and Multicast Transmissions}}

\author{\IEEEauthorblockN{Luis F. Abanto-Leon\IEEEauthorrefmark{2}, 
					 	  Matthias Hollick\IEEEauthorrefmark{2}, 
						  Bruno Clerckx\IEEEauthorrefmark{3}, 
						  Gek Hong (Allyson) Sim\IEEEauthorrefmark{2}, 
						}	
								
\IEEEauthorblockA{\IEEEauthorrefmark{2}Secure Mobile Networking (SEEMOO) Lab, Technical University of Darmstadt, Germany \\
\IEEEauthorrefmark{3}Imperial College London, United Kingdom \\
Email:
\IEEEauthorrefmark{2}\{labanto,asim,mhollick\}@seemoo.tu-darmstadt.de,
\IEEEauthorrefmark{3}b.clerckx@imperial.ac.uk
}
}

\markboth{Journal of \LaTeX\ Class Files,~Vol.~33, No.~27, June~2019}%
{Shell \MakeLowercase{\textit{et al.}}: Bare Demo of IEEEtran.cls for IEEE Communications Society Journals}

\maketitle

\begin{abstract}


This paper investigates rate-splitting (RS) precoding for non-orthogonal unicast and multicast (NOUM) transmissions using fully-digital and hybrid precoders. We study the nonconvex weighted sum-rate (WSR) maximization problem subject to a multicast requirement. We propose \texttt{FALCON}, an approach based on sequential parametric optimization, to solve the aforementioned problem. We show that \texttt{FALCON} converges to a local optimum without requiring judicious selection of an initial feasible point. Besides, we show through simulations that by leveraging RS, hybrid precoders can attain nearly the same performance as their fully-digital counterparts under certain specific settings.

\end{abstract}


\IEEEpeerreviewmaketitle

\section{Introduction} \label{introduction}
The upsurge of wireless applications and the increasing omnipresence of internet-of-things (IoT) devices are expected to exacerbate the scarcity of radio resources. Moreover, recent requirements to support multiple services simultaneously (e.g., concurrent unicast and multicast) will further aggravate the situation. This state of affairs has invigorated research on non-orthogonal unicast and multicast (NOUM) transmissions and millimeter-wave. On the one hand, NOUM has been envisaged as a recourse for improving spectral efficiency due to its capability of providing concurrent services in the same radio frequency resources. Specifically, NOUM has encountered outstanding solutions through layered-division-multiplexing (LDM) \cite{zhao2016:ldm-broadcast-unicast} and rate-splitting (RS) \cite{mao2018:rate-splitting-noma-unicast-multicast}. On the other hand, usage of the underexploited millimeter-wave spectrum is a promising solution to alleviate the scarcity of radio resources. However, the adoption of (traditional) fully-digital precoders at high frequencies is challenging owing to prohibitive fabrication costs and excessive power consumption. Consequently, the more power-efficient hybrid precoders have emerged as a functional solution to overcome such difficulties. 

\noindent \textit{\textbf{Motivation:}} \emph{Inspired by } \emph{the exceptional spectral efficiency achieved by fully-digital precoders in RS-NOUM \cite{mao2018:rate-splitting-noma-unicast-multicast, mao2019:rate-splitting-noma-spectral-energy-efficiency}, and } \emph{the predominant use of hybrid precoders in millimeter-wave communications, we investigate the weighted sum-rate (WSR) maximization problem with both fully-digital and hybrid precoders in millimeter-wave RS-NOUM systems.} Further, the body of work on WSR maximization with RS precoding leverages weighted minimum mean square error (\texttt{WMMSE}) \cite{mao2018:rate-splitting-noma-unicast-multicast, joudeh2016:rate-splitting-robust-multiuser-transmission,  joudeh2016:rate-splitting-sum-rate-maximization-partial-csit, christensen2008:weighted-sum-rate-maximization-wmmse-beamforming} as a prevailing means of solution. In this paper, we propose \texttt{FALCON}, which is based on sequential parametric optimization and compared to \texttt{WMMSE}. Our contributions are: 

\begin{itemize} [leftmargin=0.4cm]
	\item We propose \texttt{FALCON}, a fast convergence algorithm for solving the nonconvex WSR maximization problem. Our approach leverages semidefinite programming (SDP) and successive convex approximations (SCA). Specifically, we exploit sub-level/super-level sets and establish parametric convex upper bounds that can be contracted iteratively \cite{beck2010:sequential-parametric-convex-approximation-nonconvex-problems}. We show that \texttt{FALCON} converges to a local optimum of the nonconvex WSR maximization problem. Further, \texttt{FALCON} does not rely on judicious selection of initial feasible points (as \texttt{WMMSE} does) in order to guarantee high performance.
	
	\item We use \texttt{FALCON} and \texttt{WMMSE} with both RS-NOUM and LDM-NOUM systems. We realize that the former outperforms the latter due to the capability of RS to partially decode interference and partially treat remaining interference as noise. We compare RS-NOUM against the optimal dirty paper coding, thus revealing a small optimality gap between the two schemes.
	
	\item We find that, under specific settings, quantizing the hybrid precoder phase shifts with 4 bits is sufficient to achieve the same performance as fully-digital implementations. Further, in scenarios wherein users have channels with a low degree of correlation, such performance can be attained with 2 bits.
	
	\item We show, through simulations, that \texttt{FALCON} is less prone than \texttt{WMMSE} to return infeasible solutions due to its non-dependency on initial feasible points. Nevertheless, \texttt{FALCON} has higher computational complexity per iteration.
	
	
\end{itemize}

\section{Related Work} 

NOUM has been studied under LDM in \cite{zhao2016:ldm-broadcast-unicast, zhao2019:ldm-broadcast-unicast, liu2017:joint-multicast-unicast-beamforming-miso, abanto2020:fairness-hybrid-precoding-unicast-multicast,  li2019:energy-efficient-noma-multicast-unicast, tervo2017:energy-efficient-unicast-multicast-beamforming, chen2017:backhaul-constrained-beamforming-multicast-unicast, chen2018:bs-clustering-beamforming-ldm-backhaul}. Specifically, \cite{zhao2016:ldm-broadcast-unicast, zhao2019:ldm-broadcast-unicast, liu2017:joint-multicast-unicast-beamforming-miso} investigate the design of fully-digital precoders for transmit power minimization with unicast and multicast quality-of-service (QoS) requirements. A fairness-aware hybrid precoder is proposed in \cite{abanto2020:fairness-hybrid-precoding-unicast-multicast, abanto2021:beamwave-cross-layer-beamforming-scheduling-superimposed-transmissions-industrial-iot-mmwave-networks}. Energy efficiency with fully-digital precoders is investigated in \cite{li2019:energy-efficient-noma-multicast-unicast, tervo2017:energy-efficient-unicast-multicast-beamforming} whereas a similar setting with backhaul constraints is considered in \cite{chen2017:backhaul-constrained-beamforming-multicast-unicast, chen2018:bs-clustering-beamforming-ldm-backhaul}. On the other hand, NOUM with RS has only been researched with fully-digital precoders in \cite{mao2018:rate-splitting-noma-unicast-multicast, mao2019:rate-splitting-noma-spectral-energy-efficiency}, wherein the \texttt{WMMSE} approach is proposed for WSR maximization. Relevant literature on other aspects of RS includes \cite{joudeh2016:rate-splitting-sum-rate-maximization-partial-csit, joudeh2016:rate-splitting-robust-multiuser-transmission, tervo2018:rate-splitting-multigroup-multicasting-antenna-selection, chen2018:rate-splitting-multigroup-multicasting-multicarrier-systems, su2019:rate-splitting-swipt-imperfect-csit, mao2019:rate-splitting-generalization-sdma-noma, dai2017:multiuser-mmwave-beamforming-quantized-csit}. \emph{Although both LDM and RS are power-domain non-orthogonal schemes, the superiority of RS resides in its capability to partially decode interference. This is achieved by embedding fragments of unicast information in the multicast signal, that the receivers can decode and remove.}



\section{System Model: RS for NOUM Transmissions} \label{system_model}
We consider a system where a next-generation Node B (gNodeB) serves $ K $ single-antenna users indexed by $ \mathcal{K} = \left\lbrace 1, \dots, K \right\rbrace $. The gNodeB is equipped with a hybrid transmitter composed of $ N_\mathrm{tx} $ transmit antennas and $ N^\mathrm{RF}_\mathrm{tx} $ radio frequency (RF) chains, where $ K \leq N^\mathrm{RF}_\mathrm{tx} \leq N_\mathrm{tx} $. Without loss of generality, we assume $ N^\mathrm{RF}_\mathrm{tx} = K $. The gNodeB transmits in a non-orthogonal manner a multicast message $ W^{(\mathrm{m})} $ (intended to all the users) and $ K $ unicast messages $ W^{(\mathrm{u})}_1, \dots, W^{(\mathrm{u})}_K $ (each targeting a particular user). Every unicast message is decomposed into two components as $ W^{(\mathrm{u})}_k \triangleq \left( W^{(\mathrm{u,c})}_k, W^{(\mathrm{u,p})}_k \right) $, where $ W^{(\mathrm{u,c})}_k $ and $ W^{(\mathrm{u,p})}_k $ are referred as the common and private parts, respectively. The multicast message along with the unicast common parts are jointly encoded into a common macro-stream as $ \left\lbrace W^{(\mathrm{u,c})}_1, \dots, W^{(\mathrm{u,c})}_K, W^{(\mathrm{m})} \right\rbrace \mapsto \widetilde{\mathbf{z}} = \left[ \widetilde{z}_1, \widetilde{z}_2, \dots \right]^T $. The unicast private parts are encoded into independent streams as $ W^{(\mathrm{u,p})}_k \mapsto \widetilde{\mathbf{s}}_k = \left[ \widetilde{s}_{k,1}, \widetilde{s}_{k,2}, \dots \right]^T $, $ \forall k \in \mathcal{K} $. The encoded streams are processed by the baseband digital precoder $ \left[ \mathbf{B} \vert \mathbf{m} \right] = \left[ \mathbf{b}-1, \dots, \mathbf{b}_K, \mathbf{m} \right] \in \mathbb{C}^{K \times (K + 1)} $ of the hybrid transmitter. Let $ z $ and $ \mathbf{s} = \left[ s_1, \dots, s_K \right]^T $ denote the instantaneous symbol and vector symbol of the common macro-stream and private streams, respectively, such that $ \mathbb{E} \left\lbrace \left[ \mathbf{s}^T \vert z \right]^H \left[ \mathbf{s}^T \vert z \right] \right\rbrace = \mathbf{I} $. Thus, the downlink signal is $ \mathbf{x} = \mathbf{F} \left[ \mathbf{B} \vert \mathbf{m} \right] \left[ \mathbf{s}^T \vert z \right]^T \in \mathbb{C}^{N_\mathrm{tx} \times 1} $, where $\mathbf{F} = \left[ \mathbf{f}_1, \dots, \mathbf{f}_K \right] \in \mathbb{C}^{N_\mathrm{tx} \times K} $ represents the RF analog precoder of the hybrid transmitter. Under flat fading, the signal received by the $ k $-th user is
\begin{align} \label{e1}
		y_k = \underbrace{\mathbf{h}^H_k \mathbf{F} \mathbf{m} z}
		_{\substack{\text{common signal} \\ y^{(\mathrm{c})}_k}} + \underbrace{\mathbf{h}^H_k \mathbf{F} \mathbf{b}_k s_k}
		_{\substack{\text{private signal for user } k \\ y^{(\mathrm{p})}_k}} + \underbrace{\mathbf{h}^H_k \mathbf{F} \textstyle \sum \nolimits_{j \neq k}{\mathbf{b}_j s_j}}_{\substack{\text{interference at user } k \\ y^{(\mathrm{i})}_k}} + \underbrace{n_k}_{\text{noise}}, 
\end{align}
where $ n_k \sim \mathcal{CN} \left( 0, \sigma^2 \right) $ denotes additive white Gaussian noise and $\mathbf{h}_k \in {\mathbb{C}}^{N_\mathrm{tx} \times 1}$ represents the channel between the gNodeB and the $ k $-th user. Because the users have a single antenna, they are not capable of performing spatial demultiplexing to separate the two data streams. In order to distinguish and decode both signals, the users rely on successive interference cancellation (SIC), which consists in detecting and decoding one signal after the other, as explained in the following. First, each user $ k $ decodes the common macro-stream symbol $ z $, in term $ y^{(\mathrm{c})}_k $, by treating the rest of signals as noise. Subsequently, the common signal $ y^{(\mathrm{c})}_k $ is reconstructed and subtracted from $ y_k $. At this point, the remaining byproduct $ \tilde{y}_k = y_k - y^{(\mathrm{c})}_k $ consists solely of private unicast components $ \left\lbrace y^{(\mathrm{p})}_k, y^{(\mathrm{i})}_k \right\rbrace $ and noise $ n_k $, from where user $ k $ decodes symbol $ s_k $. Thus, the signal--to--interference--plus--noise ratio (SINR) of the common macro-stream and private streams are denoted by $ \mathrm{SINR}^{(\mathrm{c})}_k $ and $ \mathrm{SINR}^{(\mathrm{p})}_k $, respectively.
\resizebox{1.01\columnwidth}{!}{
\begin{minipage}{1.11\columnwidth}
	\begin{equation} \nonumber
		\mathrm{SINR}^{(\mathrm{c})}_k = \frac{\left| \mathbf{h}^H_k \mathbf{F} \mathbf{m} \right|^2 }
						   {
				                  \sum_{j \in \mathcal{K}} \left| \mathbf{h}^H_k \mathbf{F} \mathbf{b}_j \right|^2 + {\sigma}^2 
			           	   } ~
		\mathrm{SINR}^{(\mathrm{p})}_k = \frac{\left| \mathbf{h}^H_k \mathbf{F} \mathbf{b}_k \right|^2 }
				   { 
				   \sum_{j \neq k} \left| \mathbf{h}^H_k  \mathbf{F} {\mathbf{b}_j} \right|^2 + \sigma^2
				   }
	\end{equation}
\end{minipage}
} 

\noindent\textbf{RS principle:} Based on the expressions above, the achievable rates are $ R^{(\mathrm{c})}_k = \log_2 \left( 1 + \mathrm{SINR}^{(\mathrm{c})}_k \right) $ and $ R^{(\mathrm{p})}_k = \log_2 \left( 1 + \mathrm{SINR}^{(\mathrm{p})}_k \right) $. We define $ \bar{R}^{(\mathrm{c})} $ as the maximal rate at which all users can successfully decode the common symbol $ z $. Thus, the common macro-stream is to be encoded at a rate $ \bar{R}^{(\mathrm{c})} \leq R^{(\mathrm{c})}_\mathrm{min} $, where $ R^{(\mathrm{c})}_\mathrm{min} = \min_{k \in \mathcal{K}} \left\lbrace R^{(\mathrm{c})}_1, \dots, R^{(\mathrm{c})}_K \right\rbrace $. Since $ z $ results from jointly encoding multiple messages, let $ C_0 $ denote the fraction of $ \bar{R}^{(\mathrm{c})} $ conveying the multicast message $ W^{(\mathrm{m})} $, and $ C_k $ the fraction of $ \bar{R}^{(\mathrm{c})} $ transmitting the unicast common part $ W^{(\mathrm{u,c})}_k $, subject to $ C_0 + \sum_{k \in \mathcal{K}} C_k = \bar{R}^{(\mathrm{c})} $. Upon decoding the streams $ \widetilde{\mathbf{z}} $ and $ \widetilde{\mathbf{s}}_k $, user $ k $ acquires $ W^{(\mathrm{u,c})}_1, \dots, W^{(\mathrm{u,c})}_K, W^{(\mathrm{m})} $ and $ W^{(\mathrm{u,p})}_k $, from where the unicast message $ W^{(\mathrm{u})}_k $ can be reassembled. The common parts $ W^{(\mathrm{u,c})}_{j \neq k} $ decoded by user $ k $ are used for interference decoding, thus improving the SINR. The aggregate unicast rate of user $ k $ is $ R^{(\mathrm{u})}_k = C_k + \log_2 \left( 1 + \mathrm{SINR}^{(\mathrm{p})}_k \right) $.

\section{Problem Formulation} \label{problem_formulation}
The QoS-constrained WSR maximization problem is
\begin{subequations} \label{e3}
	\begin{align}
		\mathcal{P}: & \max_{
				\substack{
							C_0, C_k, \mathbf{F}, \mathbf{b}_k , \mathbf{m}
						 }
			   }
		& & \textstyle \sum \nolimits_{k \in \mathcal{K}} \mu_k \left( C_k + \log_2 \left( 1 + \mathrm{SINR}^{(\mathrm{p})}_k \right) \right)  \label{e3a}
		\\
		& ~~~~~~ \mathrm{s.t.} & & \textstyle C_0 + \sum \nolimits_{j \in \mathcal{K}} C_j = \bar{R}^{(\mathrm{c})}, \label{e3b}
		\\
		& & & C_0 \geq C^\mathrm{th}_0, \label{e3c}
		\\
		& & & C_k \geq 0, \forall k \in \mathcal{K}, \label{e3d}
		\\
		& & & \textstyle \left\| \mathbf{F} \mathbf{m} \right\|^2_2 + 
                           \sum \nolimits_{k \in \mathcal{K}} \left\| \mathbf{F} \mathbf{b}_k \right\|^2_2 \leq P_\mathrm{tx}, \label{e3e}
		\\ 
		& & & \left[ \mathbf{F} \right]_{n_1,n_2} \in \mathcal{F}, n_1 \in \mathcal{N}_1, n_2 \in \mathcal{N}_2, \label{e3f}
	\end{align}
\end{subequations}
where $ \mu_k > 0 $ in (\ref{e3a}) is the weight assigned to the $ k $-th unicast rate and (\ref{e3b}) enforces the sum of \textit{rate fractions} to be $ \bar{R}^{(\mathrm{c})} $. The constraint (\ref{e3c}) imposes a minimum QoS requirement $ C^\mathrm{th}_0 $, necessary for decoding the multicast message whereas (\ref{e3d}) imposes a non-negativity restriction on the rates $ C_k $.  The constraint (\ref{e3e}) restricts the transmit power to $ P_\mathrm{tx} $ while (\ref{e3f}) enforces the limitations of analog beamforming. Concretely, every phase shift $ \left[ \mathbf{F} \right]_{n_1,n_2} $ is constrained to the feasible set $ \mathcal{F} = \left\lbrace \delta_\mathrm{tx}, \dots, \delta_\mathrm{tx} \exp \left( j \frac{2 \pi \left( L_\mathrm{tx} - 1 \right)}{L_\mathrm{tx}} \right) \right\rbrace $, where $ L_\mathrm{tx} $ is the number of allowed constant-modulus phase shifts, $ \delta_\mathrm{tx} =  \sqrt{1 / N^\mathrm{RF}_\mathrm{tx}} $,  $ n_1 \in \mathcal{N}_1 = \left\lbrace 1, \dots, N_\mathrm{tx} \right\rbrace $ and $ n_2 \in \mathcal{N}_2 = \left\lbrace 1, \dots, N^\mathrm{RF}_\mathrm{tx} \right\rbrace $.

\subsection{WSR maximization using fully-digital precoders} \label{design_fully_digital_precoder}

In this case, $ \mathbf{F} = \mathbf{I} $ thereby transforming the hybrid precoder into a fully-digital precoder, where the number of antennas and RF chains are the same. Note that this assumption removes the nonconvex constraint (\ref{e3f}). As a result, $ \mathcal{P} $ and $ \mathcal{Q}_2 $ (shown in Section \ref{design_hybrid_precoder}) become equivalent.

\subsection{WSR maximization using hybrid precoders} \label{design_hybrid_precoder}
The hybrid precoder consists of a coupled architecture between an analog component $ \mathbf{F} $ and a digital component $ \left[ \mathbf{B} \vert \mathbf{m} \right] $. This structure poses a difficulty in obtaining an optimal solution for $ \mathcal{P} $. Therefore, we design the hybrid precoder by means of two suboptimal methods that prove to have high performance when compared to the fully-digital precoder (as evidenced in Section \ref{simulation_results}). These suboptimal methods consist of two stages: in \textit{Stage 1} we design the analog precoder whereas in \textit{Stage 2} we optimize the rate-splitting and digital precoders.

\textit{\textbf{Stage 1} (Analog precoder design):} To design the RF analog precoder $ \mathbf{F} = \left[ \mathbf{f}_1, \dots, \mathbf{f}_K \right] $, two methods are devised with the goal of maximizing the effective RF-to-RF channel gain of every user. \emph{Notice that $ \mathbf{F} $ is optimized only once as it is matched to the channels $ \left\lbrace \mathbf{h}_k \right\rbrace^K_{k = 1} $, which are deemed invariant for a number of channel uses.}

\noindent\textit{Codebook-based (CB):} This method designs $ \mathbf{F} $ in a column-wise manner using a codebook $ \mathcal{V} $ by means of
\begin{align} 
\mathcal{Q}_{1,k}: \mathbf{f}_k = \argmax_{
			\substack{
						\mathbf{v} \\
				 }
  } \left| \mathbf{h}^H_k \mathbf{v} \right|^2  ~ \mathrm{s.t.} ~ \mathbf{v} \in \mathcal{V} \backslash \bar{\mathcal{V}}, \forall k \in \mathcal{K},
\end{align}
where $ \bar{\mathcal{V}} $ is formed by all the columns of $ {\mathcal{V}} $ that have already been assigned to some user. Initially, $ \bar{\mathcal{V}} = \emptyset $ and for each user the number of elements of $ \bar{\mathcal{V}} $ is increased by one.

\noindent\textit{Projection-based (PB):} This methods designs $ \mathbf{F} $ in an element-wise manner, given the set of phase shifts $ \mathcal{F} $, via
\begin{align} \label{e4}
\mathcal{Q}_{1,k}: \mathbf{f}_k = \argmax_{
			\substack{
						\mathbf{r}_k \\
				 }
  } \left| \mathbf{h}^H_k \mathbf{r}_k \right|^2 ~ \mathrm{s.t.} ~ \left[ \mathbf{r}_k \right]_n \in \mathcal{F}, n \in \mathcal{N}_1, \forall k \in \mathcal{K}.
\end{align}
First, the unconstrained version of $ \mathcal{Q}_{1,k} $ is solved, whose solution collapses to the matched filter (i.e., a vector parallel to $ \mathbf{h}_k $). Then, such vector is projected onto the set $ \mathcal{F} $, thus yielding $ \mathbf{f}_k $, i.e., $ \left[ \mathbf{f}_k \right]_n = \argmax_{\phi \in \mathcal{F}} \mathrm{Re} \left\lbrace \phi \left[ \mathbf{h}_k \right]^*_n \right\rbrace $.

\textit{\textbf{Stage 2} (Rate-splitting and digital precoder design):}
With the analog precoder $ \mathbf{F} $ designed, the problem that optimizes $ C_0 $, $ \left\lbrace C_k \right\rbrace^K_{k = 1} $, $ \left\lbrace \mathbf{b}_k \right\rbrace^K_{k = 1} $, $ \mathbf{m} $ is given by 
\resizebox{1.01\columnwidth}{!}{
\begin{minipage}{1.03\columnwidth}
\begin{subequations} \label{e5}
	\begin{align}
		\hspace{-0.25cm}
		\mathcal{Q}_2: & \max_{
				\substack{ C_0, C_k, \mathbf{b}_k, \mathbf{m} }
			   				  }
		& & \textstyle \sum \nolimits_{k \in \mathcal{K}} \mu_k \left( C_k + \log_2 \left( 1 + \mathrm{SINR}^{(\mathrm{p})}_k \right) \right) \label{e5a}
		\\
		& ~~~~~~ \mathrm{s.t.} & & \textstyle  C_0 + \sum \nolimits_{j \in \mathcal{K}} C_j \leq \log_2 \left( 1 + \mathrm{SINR}^{(\mathrm{c})}_k \right), \forall k \in \mathcal{K}, \label{e5b}
		\\
		& & &  (\ref{e3c}), (\ref{e3d}), (\ref{e3e}), \label{e5c}
	\end{align}
\end{subequations}
\end{minipage}
} 
where (\ref{e3b}) has been equivalently recast as (\ref{e5b}) since $ \bar{R}^{(\mathrm{c})} \leq R^{(\mathrm{c})}_\mathrm{min} $. This nonconvex problem has been approached via the \texttt{WMMSE} method proposed in \cite{joudeh2016:rate-splitting-sum-rate-maximization-partial-csit, mao2018:rate-splitting-noma-unicast-multicast, mao2019:rate-splitting-noma-spectral-energy-efficiency}, which is shown to converge to a local optimum.

\section{Proposed Solution: \texttt{FALCON}} \label{falcon_sdr}
To solve $ \mathcal{Q}_2 $, we propose \texttt{FALCON}, which does not require judicious selection of an initial feasible point. Our approach stems from exploiting level sets and the establishment of convex upper bounds that can be contracted iteratively. Thus, assuming that $ \mathbf{g}_k = \mathbf{F}^H \mathbf{h}_k $, we equivalent recast $ \mathcal{Q}_2 $ as 
\begin{subequations} \label{e6}
	\begin{align}
		\widetilde{\mathcal{Q}}_2: & \max_{
				\substack{
							C_0, C_k, \mathbf{b}_k, \mathbf{m} \\
							r_k, t_k, z_k, q_k
						 }
			   				}
		& & \textstyle \sum \nolimits_{ k  \in \mathcal{K} } \mu_{k} \left( C_k + \log_2 \left( r_k \right) \right) \label{e6a}
		\\
		& ~~~~~ \mathrm{s.t.} & & \left| \mathbf{g}^H_k \mathbf{b}_k \right|^2 \geq \left( r_k - 1 \right) t_k, \forall k \in \mathcal{K}, \label{e6b}
		\\
		& & & \textstyle \sum \nolimits_{j \neq k} \left| \mathbf{g}^H_k \mathbf{b}_j \right|^2 + \sigma^2 \leq t_k, \forall k \in \mathcal{K}, \label{e6c}
		\\
		& & & \textstyle C_0 + \sum \nolimits_{j  \in \mathcal{K}} C_j \leq \log_2 \left( z_k \right), \forall k \in \mathcal{K}, \label{e6d}
		\\
		& & & \left| \mathbf{g}^H_k \mathbf{m} \right|^2 \geq \left( z_k - 1 \right) q_k, \forall k \in \mathcal{K}, \label{e6e}
		\\
		& & & \textstyle \sum \nolimits_{j  \in \mathcal{K}} \left| \mathbf{g}^H_k \mathbf{b}_j \right|^2 + \sigma^2 \leq q_k, \forall k \in \mathcal{K}, \label{e6f}
		\\
		& & & r_k \geq 1, \forall k \in \mathcal{K}, \label{e6g}
		\\
		& & & z_k \geq 1, \forall k \in \mathcal{K}, \label{e6h}
		\\
		& & & (\ref{e3c}), (\ref{e3d}), (\ref{e3e}), \label{e6i}
	\end{align}
\end{subequations}
In (\ref{e6a}), $ 1 + \mathrm{SINR}^{(\mathrm{p})}_k $ is lower-bounded by $ r_k $ whereas, in (\ref{e6c}), the denominator of $ \mathrm{SINR}^{(\mathrm{p})}_k $ is upper-bounded by $ t_k $. By combining these two relations, we obtain (\ref{e6b}). The constraints (\ref{e6d}), (\ref{e6e}) and (\ref{e6f}) are obtained in a similar manner. 

\noindent{\textbf{\emph{Proposition 1}:}} The formulations $ \mathcal{Q}_2 $ and $ \widetilde{\mathcal{Q}}_2 $ are equivalent.

\noindent{\emph{Proof: We prove the equivalence by contradiction. Let $ r^\star_{k'} $, $ t^\star_{k'} $, $ z^\star_{k'} $, $ q^\star_{k'} $ denote the values of the variables for user $ k' $ at the optimum. Let us assume that at the optimum, constraint (\ref{e6c}) for user $ k' $ is inactive. Under this assumption, there must exist an strictly smaller $ \tilde{t}_{k'} < t^\star_{k'} $ for which (\ref{e6c}) attains equality. This implies that there must also exist a strictly larger $ \tilde{r}_{k'} > r^\star_{k'} $ that satisfies (\ref{e6b}), thus producing a larger objective (\ref{e6a}). This result contradicts the premise that we have obtained an optimal solution. Similar relations can be derived for (\ref{e6d}), (\ref{e6e}), (\ref{e6f}), thus corroborating the equivalence of  $ \mathcal{Q}_2 $ and $ \widetilde{\mathcal{Q}}_2 $.} $ \blacksquare $}
\resizebox{1.01\columnwidth}{!}{
\begin{minipage}{1.01\columnwidth}
\begin{subequations} \label{e7}
	\begin{align}
		\hspace{-0.3cm}
		\widehat{\mathcal{Q}}_2: & \max_{
				\substack{
							C_0, C_k, \mathbf{B}_k, \mathbf{M}, \\
							r_k, t_k, z_k, q_k
						 }
			   							  }
		& & \textstyle \sum \nolimits_{ k  \in \mathcal{K} } \mu_{k} \left( C_k + \log_2 \left( r_k \right) \right) \label{e7a}
		\\
		& ~~~~~~ \mathrm{s.t.} & & r_k t_k - t_k - \mathrm{Tr} \left( \mathbf{G}_k \mathbf{B}_k \right) \leq 0, \forall k \in \mathcal{K}, \label{e7b}
		\\
		& & & \textstyle \sum \nolimits_{j \neq k} \mathrm{Tr} \left( \mathbf{G}_k \mathbf{B}_j \right) + \sigma^2 - t_k \leq 0, \forall k \in \mathcal{K}, \label{e7c}
		\\
		& & & \textstyle C_0 + \sum \nolimits_{j  \in \mathcal{K}} C_j \leq \log_2 \left( z_k \right), \forall k \in \mathcal{K}, \label{e7d}
		\\
		& & & z_k q_k - q_k - \mathrm{Tr} \left( \mathbf{G}_k \mathbf{M} \right) \leq 0, \forall k \in \mathcal{K}, \label{e7e}
		\\
		& & & \textstyle \sum \nolimits_{j \in \mathcal{K}} \mathrm{Tr} \left( \mathbf{G}_k \mathbf{B}_j \right) + \sigma^2 - q_k \leq 0, \forall k \in \mathcal{K}, \label{e7f}
		\\
		& & & \textstyle \mathrm{Tr} \left( \mathbf{F}^H \mathbf{F} \mathbf{M} \right) + \sum \nolimits_{k  \in \mathcal{K}} \mathrm{Tr} \left( \mathbf{F}^H \mathbf{F} \mathbf{B}_k \right) \leq P_\mathrm{tx}, \label{e7g}
		\\
		& & & \mathbf{B}_k \succeq 0, \forall k \in \mathcal{K}, \label{e7h}
		\\
		& & & \mathbf{M} \succeq 0, \label{e7i}
		\\
		& & & (\ref{e3c}), (\ref{e3d}), (\ref{e6g}), (\ref{e6h}), \label{e7j}
	\end{align}
\end{subequations}
\end{minipage}
}

Via semidefinite programming, $ \widetilde{\mathcal{Q}}_2 $ is transformed into $ \widehat{\mathcal{Q}}_2 $, where $ \mathbf{G}_k = \mathbf{g}_k \mathbf{g}^H_k $, and the rank-one constraints on $ \mathbf{B}_k = \mathbf{b}_k \mathbf{b}^H_k $, $ \mathbf{M} = \mathbf{m} \mathbf{m}^H $ have been neglected. With the exception of constraints (\ref{e7b}), (\ref{e7e}), which contain quasi-concave functions of the form $ x y $, the rest of expressions in $ \widehat{\mathcal{Q}}_2 $ constitute a convex problem. In order to circumvent these constraints we resort to the inequality $ \frac{\gamma}{2} x^2  + \frac{1}{2 \gamma} y^2 \geq x y $, which arises from the arithmetic-geometric mean of $ \gamma x^2 $ and $ \frac{1}{\gamma} y^2 $, for $ \gamma > 0 $  \cite{beck2010:sequential-parametric-convex-approximation-nonconvex-problems}. Upon applying this upper estimate to (\ref{e7b}), (\ref{e7e}), we obtain $ \widebar{\mathcal{Q}}_2 $.
\resizebox{1.01\columnwidth}{!}{
\begin{minipage}{1.1\columnwidth}
\begin{subequations} \label{e8}
	\begin{align}
		\hspace{-0.3cm}
		\widebar{\mathcal{Q}}_2: & \max_{
				\substack{
							C_0, C_k, \mathbf{B}_k, \mathbf{M}, \\
							r_k, t_k, z_k, q_k, \alpha_k, \beta_k
						 }
			   				}
		& & \textstyle \sum \nolimits_{ k  \in \mathcal{K} } \mu_{k} \left( C_k + \log_2 \left( r_k \right) \right) \label{e8a}
		\\
		& ~~~~~~~~ \mathrm{s.t.} & & \frac{\alpha_k}{2} r^2_k + \frac{1}{2 \alpha_k} t^2_k - t_k - \mathrm{Tr} \left( \mathbf{G}_k \mathbf{B}_k \right) \leq 0, \forall k \in \mathcal{K}, \label{e8b}
		\\
		& & & \frac{\beta_k}{2} z^2_k + \frac{1}{2 \beta_k} q^2_k - q_k - \mathrm{Tr} \left( \mathbf{G}_k \mathbf{M} \right) \leq 0, \forall k \in \mathcal{K}, \label{e8c}
		\\
		& & & \alpha_k > 0, \beta_k > 0, \forall k \in \mathcal{K}, \label{e8d}
		\\
		& & & (\ref{e7c}), (\ref{e7d}), (\ref{e7f}), (\ref{e7g}), (\ref{e7h}), (\ref{e7i}), (\ref{e7j}), \label{e8e}
	\end{align}
\end{subequations}
\end{minipage}
}

\noindent{\textbf{\emph{Proposition 2}:}} When $ \gamma = y/x $, $ \widebar{\mathcal{Q}}_2 $ and $ \widehat{\mathcal{Q}}_2 $ are equivalent.

\noindent{\emph{Proof: The geometric mean equates the arithmetic mean when the two constituents are equal, which occurs at $ \gamma = y/x $.} $ \blacksquare $}

Note that $ \widebar{\mathcal{Q}}_2 $ is still nonconvex. However, we have removed the complicated constraints (\ref{e7b}), (\ref{e7e}) by incorporating auxiliary variables $ \alpha_k $ and $ \beta_k $. Notice that if $ \alpha_k $ and $ \beta_k $ are fixed, $ \widebar{\mathcal{Q}}_2 $ is convex. Therefore, we are in the position of tailoring an algorithm to solve $ \widebar{\mathcal{Q}}_2 $. By harnessing \emph{Proposition 2}, we propose Algorithm \ref{algorithm_1}, wherein the upper estimates of $ r_k t_k $ and $ z_k q_k $ are contracted iteratively by updating $ \alpha_k $ and $ \beta_k $ (i.e., \texttt{lines 3,4}). Although Algorithm \ref{algorithm_1} solves $ \widebar{\mathcal{Q}}_2 $, a solution to $ \widebar{\mathcal{Q}}_2 $ may not be feasible to $ \widehat{\mathcal{Q}}_2 $ or $ \widetilde{\mathcal{Q}}_2 $ due to omission of the rank-one constraints and the use of upper estimates. In the following, we clarify these aspects.

\noindent{\textbf{\emph{Proposition 3}:}} The solutions to $ \widebar{\mathcal{Q}}^{(i)}_2 $ have at most rank one.

\noindent{\emph{Proof: Let $ \widebar{\mathcal{Q}}^{(i)}_2 $ denote the $ i $-th iteration of $ \widebar{\mathcal{Q}}_2 $. Let the Lagrangian with respect to $ \mathbf{M} $ be defined as $ \mathcal{L}_\mathbf{M} = \sum_{k \in \mathcal{K}} \xi_k \left( \frac{\beta_k}{2} z^2_k + \frac{1}{2 \beta_k} q^2_k - q_k - \mathrm{Tr} \left( \mathbf{G}_k \mathbf{M} \right) \right)  + \sum_{k \in \mathcal{K}} \pi_k \big( z_k q_k - q_k - \mathrm{Tr} \left( \mathbf{G}_k \mathbf{M} \right) \big) + \epsilon \left( \textstyle \mathrm{Tr} \left( \mathbf{F}^H \mathbf{F} \mathbf{M} \right) + \sum \nolimits_{k  \in \mathcal{K}} \mathrm{Tr} \left( \mathbf{F}^H \mathbf{F} \mathbf{B}_k \right) - P_\mathrm{tx} \right) - \mathrm{Tr} \left( \mathbf{S} \mathbf{M} \right) $, where $ \xi_k \geq 0 $, $ \pi_k \geq 0 $, $ \epsilon \geq 0 $, $ \mathbf{S} \succcurlyeq \mathbf{0} $ are the dual variables. From the stationarity condition, we obtain $ \mathbf{S} = \epsilon (\mathbf{F}^H \mathbf{F}) - \sum_{k \in \mathcal{K}} (\xi_k + \pi_k) \mathbf{G}_k \succcurlyeq \mathbf{0} $. Since $ \mathbf{F}^H \mathbf{F} $ is positive definite, the equivalent relation $ \epsilon \mathbf{I} - \sum_{k \in \mathcal{K}} (\xi_k + \pi_k) (\mathbf{F}^H \mathbf{F})^{-1} \mathbf{G}_k \succcurlyeq \mathbf{0} $ holds if $ \epsilon \geq \lambda_\mathrm{max} (\mathbf{T}) $, where $ \lambda_\mathrm{max} $ is the principal eigenvalue of $ \mathbf{T} = \sum_{k \in \mathcal{K}} (\xi_k + \pi_k) (\mathbf{F}^H \mathbf{F})^{-1} \mathbf{G}_k \succcurlyeq \mathbf{0} $. When $ \epsilon > \lambda_\mathrm{max} $, the matrix $ \mathbf{S} $ is positive definite and therefore $ \mathrm{rank}(\mathbf{S}) = N^\mathrm{tx}_\mathrm{RF} $. Due to the complementary slackness condition, this implies that $ \mathbf{M} = \mathbf{0} $. However, replacing $ \mathbf{M} = \mathbf{0} $ in $ \widebar{\mathcal{Q}}^{(i)}_2 $ yields $ C_0 = 0 $, which violates (\ref{e3c}). Therefore this solution is not feasible. When $ \epsilon = \lambda_\mathrm{max} $ then $ \mathbf{S} \succcurlyeq \mathbf{0} $. Consequently, $ \mathrm{rank} (\mathbf{S}) = N^\mathrm{tx}_\mathrm{RF} -1 $ and $ \mathrm{rank} (\mathbf{M}) = 1 $, which satisfies the assertion. For the unicast precoders $ \mathbf{B}_k $, similar relations can be derived. However, $ \mathbf{B}_k = \mathbf{0} $ does not violate any constraint. In particular, $ \mathbf{B}_k = \mathbf{0} $ can be optimal for a particular instance, specifically when the user weight $ \mu_k $ is small compared to other weights. } $ \blacksquare $ }

\noindent{\textbf{\emph{Proposition 4}:}} The sequence of objective function values produced by the update method in Algorithm \ref{algorithm_1} converges.

\noindent{\emph{Proof: Let $ \Omega_i $ denote the optimum of $ \widebar{\mathcal{Q}}^{(i)}_2 $. Also, let $ f \left( \Omega_i \right) $ be the objective function of $ \widebar{\mathcal{Q}}^{(i)}_2 $ evaluated at the optimum $ \Omega_i $. Note that if we evaluate $ f \left( \Omega_{i+1} \right) $, all the constraints are still satisfied. This implies that an optimal solution to $ \widebar{\mathcal{Q}}^{(i)}_2 $ is feasible to $ \widebar{\mathcal{Q}}^{(i+1)}_2 $. Further, the update method (i.e., \texttt{line 4}) renders $ f \left( \Omega_{i+1} \right) \geq f \left( \Omega_i \right) $, thus generating a monotonically non-decreasing sequence of objective function values. Moreover, since $ \widebar{\mathcal{Q}}_2 $ is limited by a power constraint, the non-decreasing sequence will be bounded and therefore guarantees convergence.} $ \blacksquare $ }

\noindent{\textbf{\emph{Proposition 5}:}} $ \widebar{\mathcal{Q}}_2 $ satisfies the KKT conditions of $ \mathcal{Q}_2 $.

\noindent{\emph{Proof: A solution $ \Omega_i $ to $ \widebar{\mathcal{Q}}^{(i)}_2 $ will always be feasible to $ \widehat{\mathcal{Q}}_2 $ as the inequalities of $ \widebar{\mathcal{Q}}^{(i)}_2 $ are tighter. Because a solution $ \Omega_i $ with rank-one $ \mathbf{M}^{(i)} $, $ \mathbf{B}^{(i)}_k $, $ \forall k \in \mathcal{K} $ can be found in $ \widebar{\mathcal{Q}}^{(i)}_2 $, $ \Omega_i $ is also feasible to $ \widetilde{\mathcal{Q}}_2 $  and $ \mathcal{Q}_2 $. Invoking the results in Proposition 3.2 of \cite{beck2010:sequential-parametric-convex-approximation-nonconvex-problems}, we state that the sequence of solutions $ \left\lbrace \Omega_i \right\rbrace $ converges to a regular point $ \Omega^\star_i $ that is a KKT point of $ \widetilde{\mathcal{Q}}_2 $ and $ \mathcal{Q}_2 $.} $ \blacksquare $ }

Based on the propositions above, we show that via \texttt{FALCON} the feasible set of $ \widebar{\mathcal{Q}}_2 $ converges to the feasible set of $ \mathcal{Q}_2 $. Further, by iteratively solving $ \widebar{\mathcal{Q}}^{(i)}_2 $ the solutions $ \Omega_i $ converge to a local optimum of the nonconvex problem $ \mathcal{Q}_2 $.

\setlength{\textfloatsep}{5pt}
\begin{algorithm} [!h]
	\scriptsize
	\textbf{Input:} $ \left\lbrace \mathbf{h}_k \right\rbrace^K_{k = 1} $, $ \left\lbrace \mu_k \right\rbrace^K_{k = 1} $, $ C^\mathrm{th}_0 $. \\
	\textbf{Output:} $ \mathbf{F} $, $ \left\lbrace \mathbf{b}_k \right\rbrace^K_{k = 1} $, $ \mathbf{m} $, $ C_0 $, $ \left\lbrace C_k \right\rbrace^K_{k = 1} $ \\
	\textbf{Execute:} \\ 
	\vspace{0.1cm}
		\begin{tabular}{m{0.2cm} m{7.2cm}}
			1: & Design the analog precoder $ \mathbf{F} $ by solving $ \mathcal{Q}_{1,k} $, $ \forall k \in \mathcal{K} $. \\
			2: & Initialize $ \alpha^{(i)}_k = 1 $ and $ \beta^{(i)}_k = 1 $, $ i = 0 $, $ \forall k \in \mathcal{K} $. \\
			\textbf{repeat} \\
			3: & Solve $ \widebar{\mathcal{Q}}^{(i)}_2 $ employing $ \alpha^{(i)}_k $ and $ \beta^{(i)}_k $. \\
			4: & Update $ \alpha^{(i+1)}_k = t^{(i)}_k / r^{(i)}_k $, $ \beta^{(i+1)}_k = q^{(i)}_k / z^{(i)}_k $, $ \forall k \in \mathcal{K} $. \\
			5: & Update $ i = i + 1 $. \\
			\textbf{until} & $ \mathrm{stop criterion} $.
		\end{tabular}
	\caption{\footnotesize Rate-splitting and precoding via \texttt{FALCON}}
	\label{algorithm_1}
\end{algorithm}

\section{Simulation Results} \label{simulation_results}
In this section, we compare \texttt{FALCON} and \texttt{WMMSE}. In the scenarios evaluated in Section \ref{simulations_feasibility_response}, Section \ref{simulations_convergence}, Section \ref{simulations_two_user_region} we employ the more versatile projection-based hybrid precoder whereas in Section \ref{simulations_hybrid_precoder_designs} we compare the two types of hybrid precoders. Throughout the simulations, we have considered $ P_\mathrm{tx} = 50 $ dBm and $ \sigma^2 = 30 $ dBm as in \cite{mao2018:rate-splitting-noma-unicast-multicast}. In addition, for all scenarios involving different precoders and techniques, we have used the same stopping criterion (see Algorithm \ref{algorithm_1}). Specifically, the techniques are executed for a maximum of $ N_\mathrm{iter} = 60 $ iterations or until an increment of less than $ \epsilon = 0.0001 $ is attained (by the objective function). 

\subsection{Feasibility response} \label{simulations_feasibility_response}

\begin{table}[!t]
	\vspace{2mm}
	\fontsize{5}{6}\selectfont
	\centering
	\setlength\tabcolsep{1.5pt}
	\renewcommand{\arraystretch}{0.25}
	\caption{Feasibility response of fully-digital precoders.}
	\begin{tabular}{c c c c c c c c c c c c c c c c c c}
		\toprule
		\multirow{3}{*}{\rotatebox{90}{\kern-2.25em Case}} & 
		\multirow{1}{*}{Parameters} & 
		\multirow{2}{*}{\texttt{FALCON}} &
		\multicolumn{15}{c}{\texttt{WMMSE} (\%)} \\ 
		\cmidrule{4-18} 
		& configuration &  & \multicolumn{3}{c}{\scalebox{.65}{$ P^{(0)}_\mathrm{m} = 0.70 P_\mathrm{tx} $}} & \multicolumn{3}{c}{\scalebox{.65}{$ P^{(0)}_\mathrm{m} = 0.80 P_\mathrm{tx} $}} & \multicolumn{3}{c}{\scalebox{.65}{$ P^{(0)}_\mathrm{m} = 0.90 P_\mathrm{tx} $}} & \multicolumn{3}{c}{\scalebox{.65}{$ P^{(0)}_\mathrm{m} = 0.95 P_\mathrm{tx} $}} & \multicolumn{3}{c}{\scalebox{.65}{$ P^{(0)}_\mathrm{m} = 0.99 P_\mathrm{tx} $}} \\
		\cmidrule(lr){3-3}  
		\cmidrule(lr){4-6}  
		\cmidrule(lr){7-9} 
		\cmidrule(lr){10-12} 
		\cmidrule(lr){13-15} 
		\cmidrule(lr){16-18} 
		& \scalebox{.65}{$ { \left[ N_\mathrm{tx}, K, C^\mathrm{th}_0 \right]} $}  & (\%) & \texttt{MRT} & \texttt{ZF} & \texttt{SLNR} & \texttt{MRT} & \texttt{ZF} & \texttt{SLNR} & \texttt{MRT} & \texttt{ZF} & \texttt{SLNR} & \texttt{MRT} & \texttt{ZF} & \texttt{SLNR} & \texttt{MRT} & \texttt{ZF} & \texttt{SLNR} \\
		\midrule
		
		A & $ {\scriptscriptstyle \left[ 4, 2, 2.5 \right]} $ & \textbf{100} & 38 & 55 & 54 & 63 & 63 & 63 & 75 & 77 & 73 & 83 & 83 & 78 & \textbf{\underline{84}} & 83 & 80 \\
		\midrule
		
		B & $ {\scriptscriptstyle \left[ 6, 3, 2.0 \right]} $ & \textbf{100} & 36 & 49 & 68 & 51 & 63 & 70 & 70 & 80 & 77 & 81 & 83 & 81 & \textbf{\underline{87}} & 87 & 83 \\
		\midrule	
		
		C & $ {\scriptscriptstyle \left[ 8, 4, 1.5 \right]} $ &  \textbf{96} & 53 & 67 & 86 & 69 & 73 & 89 & 80 & 85 & 93 & 83 & 85 & 93 & 91 & 92 & \textbf{\underline{94}} \\

		\bottomrule
	\end{tabular}
	\label{t1}
\end{table}
\begin{table}[!t]
	\fontsize{5}{6}\selectfont
	\centering
	\setlength\tabcolsep{1.5pt}
	\renewcommand{\arraystretch}{0.25}
	\caption{Feasibility response of hybrid precoders.}
	\begin{tabular}{c c c c c c c c c c c c c c c c c c}
		\toprule
		\multirow{3}{*}{\rotatebox{90}{\kern-2.25em Case}} & 
		\multirow{1}{*}{Parameters} & 
		\multirow{2}{*}{\texttt{FALCON}} &
		\multicolumn{15}{c}{\texttt{WMMSE} (\%)} \\ 
		\cmidrule{4-18} 
		& configuration &  & \multicolumn{3}{c}{\scalebox{.65}{$ P^{(0)}_\mathrm{m} = 0.70 P_\mathrm{tx} $}} & \multicolumn{3}{c}{\scalebox{.65}{$ P^{(0)}_\mathrm{m} = 0.80 P_\mathrm{tx} $}} & \multicolumn{3}{c}{\scalebox{.65}{$ P^{(0)}_\mathrm{m} = 0.90 P_\mathrm{tx} $}} & \multicolumn{3}{c}{\scalebox{.65}{$ P^{(0)}_\mathrm{m} = 0.95 P_\mathrm{tx} $}} & \multicolumn{3}{c}{\scalebox{.65}{$ P^{(0)}_\mathrm{m} = 0.99 P_\mathrm{tx} $}} \\
		\cmidrule(lr){3-3}  
		\cmidrule(lr){4-6}  
		\cmidrule(lr){7-9} 
		\cmidrule(lr){10-12} 
		\cmidrule(lr){13-15} 
		\cmidrule(lr){16-18} 
		& \scalebox{.65}{$ { \left[ N_\mathrm{tx}, K, C^\mathrm{th}_0 \right]} $}  & (\%) & \texttt{MRT} & \texttt{ZF} & \texttt{SLNR} & \texttt{MRT} & \texttt{ZF} & \texttt{SLNR} & \texttt{MRT} & \texttt{ZF} & \texttt{SLNR} & \texttt{MRT} & \texttt{ZF} & \texttt{SLNR} & \texttt{MRT} & \texttt{ZF} & \texttt{SLNR} \\
		\midrule
		
		A & $ {\scriptscriptstyle \left[ 4, 2, 2.5 \right]} $ & \textbf{100} & 54 & 58 & 69 & 82 & 77 & 68 & 82 & 82 & 78 & 87 & 87 & 84 & \textbf{\underline{87}} & 87 & 87 \\
		\midrule
		
		B & $ {\scriptscriptstyle \left[ 6, 3, 2.0 \right]} $ & \textbf{100} & 63 & 54 & 62 & 70 & 72 & 67 & 79 & 79 & 76 & 84 & 83 & 81 & \textbf{\underline{88}} & 87 & 85 \\
		\midrule	
		
		C & $ {\scriptscriptstyle \left[ 8, 4, 1.5 \right]} $  & \textbf{96} & 74 & 81 & 79 & 85 & 85 & 86 & 93 & 94 & 91 & 93 & 95 & 94 & \textbf{\underline{96}} & 96 & 96 \\

		\bottomrule
	\end{tabular}
	\label{t2}
\end{table}
\begin{figure}[!t]
\begin{tikzpicture}
  	\begin{groupplot}
  	[
	  	group style = {group size = 2 by 1, horizontal sep = 0.6cm}, 
	  	height = 5cm, width = 5.1cm
  	]
    \nextgroupplot
    [
    	xmin = 1,
    	xmax = 50,
    	ymin = 13,
    	ymax = 25.5,
    	xlabel = {(a) \footnotesize Iteration index},
    	ylabel = {\footnotesize WSR [bps/Hz]},
    	x tick label style = {font = \fontsize{7}{6}\selectfont},
    	y tick label style = {font = \fontsize{7}{6}\selectfont},
    	xtick = {10, 20, 30, 40, 50},
    	ytick = {6, 15, 24},
    	xtick pos = left,
    	ytick pos = left,
    	legend style = {row sep = 0.01cm},
    	legend cell align = {left},
    	legend columns = 1,
    	legend pos = north east,
    	legend style = {at = {(0.44, 0.10)}, anchor = south west, font = \fontsize{5}{4}\selectfont, text depth = .ex, fill = none, draw = none},
 	]
   	\addlegendimage{color = red, mark = square*, mark options = {scale = 0.8, fill = red, solid}, line width = 0.3pt, only marks} \addlegendentry{\texttt{FALCON}}
   	
   	\addlegendimage{color = green!80!black, mark = square*, mark options = {scale = 0.8, fill = green!80!black, solid}, line width = 0.3pt, only marks}	\addlegendentry{\texttt{WMMSE|MRT|0.90}}
   	
   	\addlegendimage{color = cyan, mark = square*, mark options = {scale = 0.8, fill = cyan, solid}, line width = 0.3pt, only marks} \addlegendentry{\texttt{WMMSE|ZF|0.90}}
   	
   	\addlegendimage{color = magenta!50!white, mark = square*, mark options = {scale = 0.8, fill = magenta!50!white, solid}, line width = 0.3pt, only marks} \addlegendentry{\texttt{WMMSE|SLNR|0.90}}

   	\addplot[color = red, mark = none, line width = 0.8pt] table {data/convergence/a/FALCON_Digital_Unicast.txt};
   	
   	\addplot[color = green!80!black, mark = none, line width = 0.8pt] table {data/convergence/a/WMMSE_Digital_MRT_with_SVD_Unicast.txt};
   	
   	\addplot[color = cyan, mark = none, line width = 0.8pt] table {data/convergence/a/WMMSE_Digital_ZF_with_SVD_Unicast.txt};
   		
   	\addplot[color = magenta!50!white, mark = none, line width = 0.8pt] table {data/convergence/a/WMMSE_Digital_SLNR_with_SVD_Unicast.txt};

   	\addplot[color = red, mark = none, line width = 0.8pt, densely dotted] table {data/convergence/a/FALCON_Hybrid_Unicast.txt};
   		
   	\addplot[color = green!80!black, mark = none, line width = 0.8pt, densely dotted] table {data/convergence/a/WMMSE_Hybrid_MRT_with_SVD_Unicast.txt};		
   	
   	\addplot[color = cyan, mark = none, line width = 0.8pt, densely dotted] table {data/convergence/a/WMMSE_Hybrid_ZF_with_SVD_Unicast.txt};
   		
   	\addplot[color = magenta!50!white, mark = none, line width = 0.8pt, densely dotted] table {data/convergence/a/WMMSE_Hybrid_SLNR_with_SVD_Unicast.txt};	
   	
   	
   	\draw[rotate around = {90:(45,24.02)}, color = gray] (45,24.02) ellipse(0.1cm and 0.05cm);
   	\node[color = gray, anchor = west, right] at (25, 24.9) {\tiny fully-digital precoder};
   	
   	\draw[rotate around = {90:(45,22.60)}, color = gray] (45,22.60) ellipse(0.1cm and 0.05cm);
   	\node[color = gray, anchor = west, right] at (30, 21.6) {\tiny hybrid precoder};
	
    \nextgroupplot
    [
    	xmin = 0,
    	xmax = 50,
    	ymin = 19.9,
    	ymax = 24.1,
    	xlabel = {(b) \footnotesize Iteration index},
    	x tick label style = {font = \fontsize{7}{6}\selectfont},
    	y tick label style = {font = \fontsize{7}{6}\selectfont},
    	xtick = {10, 20, 30, 40, 50, 60},
    	ytick = {20, 22, 24},
    	xtick pos = left,
    	ytick pos = left,
    	legend style = {row sep = 0.00cm},
    	legend cell align = {left},
    	legend columns = 1,
    	legend pos = north east,
    	legend style = {at = {(0.47, 0.00)}, anchor = south west, font = \fontsize{5}{4}\selectfont, text depth = .ex, fill = none, draw = none}
    ]
    
   	\addlegendimage{color = red, mark = square*, mark options = {scale = 0.8, fill = red, solid}, line width = 0.3pt, only marks}
   	\addlegendentry{\texttt{FALCON}}
   	
   	\addlegendimage{color = orange!70!white, mark = square*, mark options = {scale = 0.8, fill = orange!70!white, solid}, line width = 0.3pt, only marks} 
   	\addlegendentry{\texttt{WMMSE|MRT|0.99}}
   	
   	\addlegendimage{color = magenta!50!white, mark = square*, mark options = {scale = 0.8, fill = magenta!50!white, solid}, line width = 0.3pt, only marks} 
   	\addlegendentry{\texttt{WMMSE|MRT|0.95}}
   	
   	\addlegendimage{color = green!80!black, mark = square*, mark options = {scale = 0.8, fill = green!80!black, solid}, line width = 0.3pt, only marks} 
   	\addlegendentry{\texttt{WMMSE|MRT|0.90}}
   	
   		
   	\addlegendimage{color = blue!50!white, mark = square*, mark options = {scale = 0.8, fill = blue!50!white, solid}, line width = 0.3pt, only marks} 
   	\addlegendentry{\texttt{WMMSE|MRT|0.80}}
   	
   	\addlegendimage{color = black!80!white, mark = square*, mark options = {scale = 0.8, fill = black!80!white, solid}, line width = 0.3pt, only marks} 
   	\addlegendentry{\texttt{WMMSE|MRT|0.70}}

   	\addplot[color = red, mark = none, line width = 0.8pt] table {data/convergence/b/FALCON_Digital_Unicast.txt};
   	
   	\addplot[color = orange!70!white, mark = none, line width = 0.8pt] table {data/convergence/b/WMMSE_Digital_MRT_with_SVD_Unicast_0.99.txt};

   	\addplot[color = magenta!50!white, mark = none, line width = 0.8pt] table {data/convergence/b/WMMSE_Digital_MRT_with_SVD_Unicast_0.95.txt};

   	\addplot[color = green!80!black, mark = none, line width = 0.8pt] table {data/convergence/b/WMMSE_Digital_MRT_with_SVD_Unicast_0.90.txt};

   	
   	\addplot[color = blue!50!white, mark = none, line width = 0.8pt] table {data/convergence/b/WMMSE_Digital_MRT_with_SVD_Unicast_0.80.txt};
   	
   	\addplot[color = black!80!white, mark = none, line width = 0.8pt] table {data/convergence/b/WMMSE_Digital_MRT_with_SVD_Unicast_0.70.txt};
   	   	
  \end{groupplot}
\end{tikzpicture}
\caption{Convergence of \texttt{FALCON} and \texttt{WMMSE}.}
\label{fig_convergence}
\end{figure}
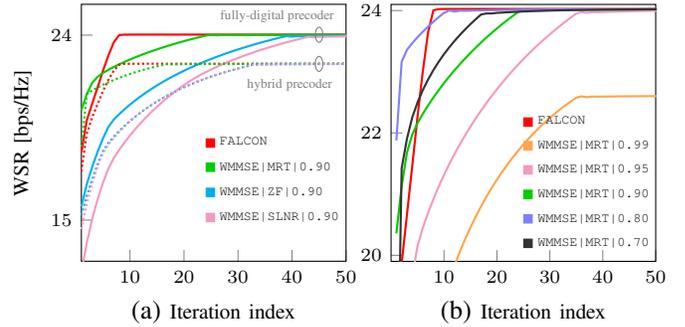

We evaluate the performance of \texttt{FALCON} and \texttt{WMMSE} with RS-NOUM, in terms of the feasible solutions count. We examine different configurations of $ \left[ N_\mathrm{tx}, K, C^\mathrm{th}_0 \right] $ assuming equal weights $ \mu_1 = \dots = \mu_K = 1 $ with fully-digital and hybrid precoders. We employ the geometric Saleh-Valenzuela channel model \cite{rappaport2015:millimeter-wave-wireless-communications}.
Since \texttt{WMMSE} requires an initial feasible point, we assess three types of initialization methods. The initial multicast precoder $ \mathbf{m}^{(0)} $ is obtained via singular value decomposition (SVD) of the aggregate channel \cite{mao2019:rate-splitting-noma-spectral-energy-efficiency, mao2019:rate-splitting-generalization-sdma-noma} whereas the initial unicast precoders $ \mathbf{b}^{(0)}_k $ are the MRT, ZF or SLNR precoding vectors. The initial multicast and unicast powers (for the initial feasible point) are computed as $ P^{(0)}_\mathrm{m} = \left\| \mathbf{F} \mathbf{m}^{(0)} \right\|^2_2 $ and $ P^{(0)}_{\mathrm{u},k} = \left\| \mathbf{F} \mathbf{b}^{(0)}_k \right\|^2_2 = \frac{P_\mathrm{tx} - P^{(0)}_\mathrm{m}}{K} $, $ \forall k \in \mathcal{K} $. We evaluate several values of $ P^{(0)} = \left\lbrace 0.70 P_\mathrm{tx}, 0.80 P_\mathrm{tx}, 0.90 P_\mathrm{tx}, 0.95 P_\mathrm{tx}, 0.99 P_\mathrm{tx} \right\rbrace $ with $ L_\mathrm{tx} = 16 $ different phase shifts for the projection-based hybrid precoder. Table \ref{t1} and Table \ref{t2} show the results for the fully-digital and hybrid precoders. Note that \texttt{FALCON} is superior to \texttt{WMSSE} in delivering a larger number of feasible solutions without requiring complicated initialization. Throughout the different configurations, \texttt{FALCON} returns a feasible solution in at least $ 96 \% $ of the cases whereas the performance of \texttt{WMMSE} is inferior. In the \texttt{WMMSE} case, the same value of $ P^{(0)} $ leads to different feasibility responses for distinct $ \left[ N_\mathrm{tx}, K, C^\mathrm{th}_0 \right] $. In certain cases, the hybrid precoder attains more feasible solutions that the fully-digital precoder. This occurs due to the strong dependence of \texttt{WMMSE} on the initial feasible point, which is also influenced by the analog precoder $ \mathbf{F} $. Nevertheless, this outcome does not imply that both precoders provide the same rate value, as clarified in the next scenario.

\begin{figure*}[!t]
\begin{minipage}[c]{\textwidth}
\begin{tikzpicture}[spy using outlines={circle, magnification=3, connect spies}]	
  \begin{groupplot}
  	[
	  	group style = {group size = 2 by 2, horizontal sep = 1.1cm, vertical sep = 0.5cm}, 
	  	height = 5.8cm, width = 9.3cm
  	]
    \nextgroupplot
    [
	    xmin = 0,
		xmax = 9, 
	    ymin = 0, 
	    ymax = 9,
	    ylabel = {\footnotesize $ R^{(\mathrm{u})}_2 $ [bps/Hz] },
	 	xtick = {0, 1, ..., 9},
	 	ytick = {0, 1, ..., 9},
	 	x tick label style = {font = \fontsize{7}{6}\selectfont},
	 	y tick label style = {font = \fontsize{7}{6}\selectfont},
	 	xtick pos = left,
	 	ytick pos = left,
	 	legend style = {row sep = 0.01cm},
	 	legend style = {column sep = 0.16cm},
    	legend cell align = {left},
    	legend columns = 11,
    	legend pos = north east,
    	legend style = {at = {(-0.05, 1.05)}, anchor = south west, font = \fontsize{7}{6}\selectfont, text depth = .ex, fill = none},
 	]
 	
 	\addlegendimage{color = black, mark = square*, mark options = {scale = 1, fill = black, solid}, line width = 0.3pt, smooth, only marks} 
 	\addlegendentry{\texttt{DPC}}
 	
   	\addlegendimage{color = red, mark = square*, mark options = {scale = 1, fill = red, solid}, line width = 0.3pt, smooth, only marks} 
   	\addlegendentry{\texttt{RS|FD}}
   	
   	\addlegendimage{color = red!50!white, mark = square*, mark options = {scale = 1, fill = red!50!white, solid}, line width = 0.3pt, smooth, only marks} 
   	\addlegendentry{\texttt{LDM|FD}}
   	
   	\addlegendimage{color = yellow, mark = square*, mark options = {scale = 1, fill = yellow, solid}, line width = 0.3pt, smooth, only marks}	\addlegendentry{\texttt{RS|H|16}}
   	
   	\addlegendimage{color = yellow!50!white, mark = square*, mark options = {scale = 1, fill = yellow!50!white, solid}, line width = 0.3pt, smooth, only marks}	\addlegendentry{\texttt{LDM|H|16}}
   	
   	\addlegendimage{color = green, mark = square*, mark options = {scale = 1, fill = green, solid}, line width = 0.3pt, smooth, only marks} 
   	\addlegendentry{\texttt{RS|H|8}}
   	
   	\addlegendimage{color = green!50!white, mark = square*, mark options = {scale = 1, fill = green!50!white, solid}, line width = 0.3pt, smooth, only marks} 
   	\addlegendentry{\texttt{LDM|H|8}}
   	
   	\addlegendimage{color = cyan, mark = square*, mark options = {scale = 1, fill = cyan, solid}, line width = 0.3pt, smooth, only marks} 
   	\addlegendentry{\texttt{RS|H|4}}
   	   	
   	\addlegendimage{color = cyan!50!white, mark = square*, mark options = {scale = 1, fill = cyan!50!white, solid}, line width = 0.3pt, smooth, only marks} 
   	\addlegendentry{\texttt{LDM|H|4}}
   	
   	\addlegendimage{color = magenta, mark = square*, mark options = {scale = 1, fill = magenta, solid}, line width = 0.3pt, smooth, only marks}  \addlegendentry{\texttt{RS|H|2}}

   	\addlegendimage{color = magenta!50!white, mark = square*, mark options = {scale = 1, fill = magenta!50!white, solid}, line width = 0.3pt, smooth, only marks}  \addlegendentry{\texttt{LDM|H|2}}

    \addplot[color = black, mark = none, line width = 0.5pt] table {data/rate region/S11/S11_Rate_Region_Two_Users_DPC.txt};
    
    \addplot[color = red, mark = none, line width = 1.2pt] table {data/rate region/S11/S11_Rate_Region_Two_Users_RS_SDR_Digital.txt};
    
   	\addplot[color = yellow, mark = none, line width = 0.6pt] table {data/rate region/S11/S11_Rate_Region_Two_Users_RS_SDR_Hybrid_16Ltx.txt};
  	
    \addplot[color = green, mark = none, line width = 0.6pt] table {data/rate region/S11/S11_Rate_Region_Two_Users_RS_SDR_Hybrid_8Ltx.txt};
    
   	\addplot[color = cyan, mark = none, line width = 1.2pt] table {data/rate region/S11/S11_Rate_Region_Two_Users_RS_SDR_Hybrid_4Ltx.txt};
       
   	\addplot[color = magenta, mark = none, line width = 0.6pt] table {data/rate region/S11/S11_Rate_Region_Two_Users_RS_SDR_Hybrid_2Ltx.txt};

   	\addplot[color = red!50!white, mark = none, line width = 1.2pt] table {data/rate region/S11/S11_Rate_Region_Two_Users_LDM_SDR_Digital.txt};
   	
   	\addplot[color = yellow!50!white, mark = none, line width = 0.6pt] table {data/rate region/S11/S11_Rate_Region_Two_Users_LDM_SDR_Hybrid_16Ltx.txt};   	
   	
   	\addplot[color = green!50!white, mark = none, line width = 0.6pt] table {data/rate region/S11/S11_Rate_Region_Two_Users_LDM_SDR_Hybrid_8Ltx.txt};
   	
   	\addplot[color = cyan!50!white, mark = none, line width = 1.2pt] table {data/rate region/S11/S11_Rate_Region_Two_Users_LDM_SDR_Hybrid_4Ltx.txt};
   	
  	\addplot[color = magenta!50!white, mark = none, line width = 0.6pt] table {data/rate region/S11/S11_Rate_Region_Two_Users_LDM_SDR_Hybrid_2Ltx.txt};

	\node[text width = 0.2cm, align = center, anchor = north] at (8.65, 9) {\footnotesize (a)};
	
	\node[draw, rectangle, minimum size = 0.5cm, gray] at (0.75,1.25) {$ \scriptscriptstyle \theta = \frac{\pi}{9}  $};
	
	\coordinate (spypoint0) at (axis cs:5.2,5.6);
	\coordinate (magnifyglass0) at (axis cs:1.4,4.6);
	\spy [gray, size=1.3cm, magnification=3.5] on (spypoint0) in node[fill=white] at (magnifyglass0);
	\node [below = 17pt of magnifyglass0,align=center] {\tiny \texttt{RS}};
	 
	\coordinate (spypoint1) at (axis cs:6.2,2.5);
	\coordinate (magnifyglass1) at (axis cs:3.5,2.6);
	\spy [gray, size=1.3cm, magnification=3.5] on (spypoint1) in node[fill=white] at (magnifyglass1);
	\node [below = 17pt of magnifyglass1,align=center] {\tiny \texttt{LDM}};

    \nextgroupplot
    [
	    xmin = 0,
   		xmax = 9, 
	    ymin = 0, 
	    ymax = 9,
	    ylabel = { \footnotesize $ R^{(\mathrm{u})}_2 $ [bps/Hz] },
   	 	xtick = {0, 1, ..., 9},
   	 	ytick = {0, 1, ..., 9},
   	 	x tick label style = {font = \fontsize{7}{6}\selectfont},
   	 	y tick label style = {font = \fontsize{7}{6}\selectfont},
   	 	xtick pos = left,
   	 	ytick pos = left,
   	 	legend style = {row sep = 0.01cm},
       	legend cell align = {left},
       	legend columns = 1,
       	legend pos = north east,
       	legend style = {at = {(0.46, 0.22)}, anchor = south west, font = \fontsize{5}{4}\selectfont, text depth = .ex, fill = none, draw = none},
    ]

    \addplot[color = black, mark = none, line width = 0.6pt] table {data/rate region/S12/S12_Rate_Region_Two_Users_DPC.txt};
    
    \addplot[color = red, mark = none, line width = 1.2pt] table {data/rate region/S12/S12_Rate_Region_Two_Users_RS_SDR_Digital.txt};
      	  	
    \addplot[color = green, mark = none, line width = 0.6pt] table {data/rate region/S12/S12_Rate_Region_Two_Users_RS_SDR_Hybrid_8Ltx.txt};
    
    \addplot[color = cyan, mark = none, line width = 0.6pt] table {data/rate region/S12/S12_Rate_Region_Two_Users_RS_SDR_Hybrid_4Ltx.txt};
    
  	\addplot[color = magenta, mark = none, line width = 0.6pt] table {data/rate region/S12/S12_Rate_Region_Two_Users_RS_SDR_Hybrid_2Ltx.txt};

    \addplot[color = red!50!white, mark = none, line width = 1.2pt] table {data/rate region/S12/S12_Rate_Region_Two_Users_LDM_SDR_Digital.txt};
         	  	
    \addplot[color = green!50!white, mark = none, line width = 0.6pt] table {data/rate region/S12/S12_Rate_Region_Two_Users_LDM_SDR_Hybrid_8Ltx.txt};
    
    \addplot[color = cyan!50!white, mark = none, line width = 0.6pt] table {data/rate region/S12/S12_Rate_Region_Two_Users_LDM_SDR_Hybrid_4Ltx.txt};
    
    \addplot[color = magenta!50!white, mark = none, line width = 0.6pt] table {data/rate region/S12/S12_Rate_Region_Two_Users_LDM_SDR_Hybrid_2Ltx.txt};
    
    \node[text width = 0.2cm, align = center, anchor = north] at (8.65, 9) {\footnotesize (b)};
    
    \node[draw, rectangle, minimum size = 0.5cm, gray] at (0.75,1.25) {$ \scriptscriptstyle \theta = \frac{2 \pi}{9} $};
    
    \coordinate (spypoint2) at (axis cs:7.625,4.6);
    \coordinate (magnifyglass2) at (axis cs:1.5,4.9);
    \spy [gray, size=1.3cm, magnification=4] on (spypoint2) in node[fill=white] at (magnifyglass2);
    \node [below = 17pt of magnifyglass2,align=center] {\tiny \texttt{RS}};
    
    \coordinate (spypoint3) at (axis cs:7.5,2.45);
    \coordinate (magnifyglass3) at (axis cs:4.5,2.6);
    \spy [gray, size=1.3cm, magnification=4] on (spypoint3) in node[fill=white] at (magnifyglass3);
    \node [below = 17pt of magnifyglass3,align=center] {\tiny \texttt{LDM}};

    \nextgroupplot
    [
	    xmin = 0,
   		xmax = 9, 
	    ymin = 0, 
	    ymax = 9,
	    xlabel = {\footnotesize $ R^{(\mathrm{u})}_1 $ [bps/Hz]},
   		ylabel = {\footnotesize $ R^{(\mathrm{u})}_2 $ [bps/Hz]},
   	 	xtick = {0, 1, ..., 9},
   	 	ytick = {0, 1, ..., 9},
   	 	x tick label style = {font = \fontsize{7}{6}\selectfont},
   	 	y tick label style = {font = \fontsize{7}{6}\selectfont},
   	 	xtick pos = left,
   	 	ytick pos = left,
   	 	legend style = {row sep = 0.01cm},
       	legend cell align = {left},
       	legend columns = 1,
       	legend pos = north east,
       	legend style = {at = {(0.46, 0.22)}, anchor = south west, font = \fontsize{5}{4}\selectfont, text depth = .ex, fill = none, draw = none},
    ]

    \addplot[color = black, mark = none, line width = 0.6pt] table {data/rate region/S13/S13_Rate_Region_Two_Users_DPC.txt};
    
    \addplot[color = red, mark = none, line width = 1.2pt] table {data/rate region/S13/S13_Rate_Region_Two_Users_RS_SDR_Digital.txt};
      	  	
    \addplot[color = green, mark = none, line width = 0.6pt] table {data/rate region/S13/S13_Rate_Region_Two_Users_RS_SDR_Hybrid_8Ltx.txt};
    
    \addplot[color = cyan, mark = none, line width = 0.6pt] table {data/rate region/S13/S13_Rate_Region_Two_Users_RS_SDR_Hybrid_4Ltx.txt};
    
  	\addplot[color = magenta, mark = none, line width = 0.6pt] table {data/rate region/S13/S13_Rate_Region_Two_Users_RS_SDR_Hybrid_2Ltx.txt};

    \addplot[color = red!50!white, mark = none, line width = 1.2pt] table {data/rate region/S13/S13_Rate_Region_Two_Users_LDM_SDR_Digital.txt};
         	  	
    \addplot[color = green!50!white, mark = none, line width = 0.6pt] table {data/rate region/S13/S13_Rate_Region_Two_Users_LDM_SDR_Hybrid_8Ltx.txt};
    
    \addplot[color = cyan!50!white, mark = none, line width = 0.6pt] table {data/rate region/S13/S13_Rate_Region_Two_Users_LDM_SDR_Hybrid_4Ltx.txt};
    
    \addplot[color = magenta!50!white, mark = none, line width = 0.6pt] table {data/rate region/S13/S13_Rate_Region_Two_Users_LDM_SDR_Hybrid_2Ltx.txt};
    
    \node[text width = 0.2cm, align = center, anchor = north] at (8.65, 9) {\footnotesize (c)};
    
    \node[draw, rectangle, minimum size = 0.5cm, gray] at (0.75,1.25) {$ \scriptscriptstyle \theta = \frac{3 \pi}{9} $};   
    
    \coordinate (spypoint4) at (axis cs:8,4.0);
    \coordinate (magnifyglass4) at (axis cs:5.5,2.5);
    \spy [gray, size = 0.8cm, magnification = 5] on (spypoint4) in node[fill=white] at (magnifyglass4); 
    \node [below = 10pt of magnifyglass4,align=center] {\tiny \texttt{RS}};
    
    \coordinate (spypoint5) at (axis cs:7.795,4.28);
	\coordinate (magnifyglass5) at (axis cs:4.1,5);
    \spy [gray, size = 0.8cm, magnification = 5] on (spypoint5) in node[fill=white] at (magnifyglass5); 
    \node [below = 10pt of magnifyglass5,align=center] {\tiny \texttt{LDM}};

    \nextgroupplot
    [
	    xmin = 0,
   		xmax = 9, 
	    ymin = 0, 
	    ymax = 9,
	    xlabel = {\footnotesize $ R^{(\mathrm{u})}_1 $ [bps/Hz]},
	    ylabel = { \footnotesize $ R^{(\mathrm{u})}_2 $ [bps/Hz] },
   	 	xtick = {0, 1, ..., 9},
   	 	ytick = {0, 1, ..., 9},
   	 	x tick label style = {font = \fontsize{7}{6}\selectfont},
   	 	y tick label style = {font = \fontsize{7}{6}\selectfont},
   	 	xtick pos = left,
   	 	ytick pos = left,
   	 	legend style = {row sep = 0.01cm},
       	legend cell align = {left},
       	legend columns = 1,
       	legend pos = north east,
       	legend style = {at = {(0.46, 0.22)}, anchor = south west, font = \fontsize{5}{4}\selectfont, text depth = .ex, fill = none, draw = none},
    ]

    \addplot[color = black, mark = none, line width = 0.6pt] table {data/rate region/S14/S14_Rate_Region_Two_Users_DPC.txt};
    
    \addplot[color = red, mark = none, line width = 1.2pt] table {data/rate region/S14/S14_Rate_Region_Two_Users_RS_SDR_Digital.txt};
    
    \addplot[color = cyan, mark = none, line width = 0.6pt] table {data/rate region/S14/S14_Rate_Region_Two_Users_RS_SDR_Hybrid_4Ltx.txt};
    
  	\addplot[color = magenta, mark = none, line width = 1.0pt] table {data/rate region/S14/S14_Rate_Region_Two_Users_RS_SDR_Hybrid_2Ltx.txt};

    \addplot[color = red!50!white, mark = none, line width = 0.9pt] table {data/rate region/S14/S14_Rate_Region_Two_Users_LDM_SDR_Digital.txt};
         	  	
    \addplot[color = cyan!50!white, mark = none, line width = 0.6pt] table {data/rate region/S14/S14_Rate_Region_Two_Users_LDM_SDR_Hybrid_4Ltx.txt};
    
    \addplot[color = magenta!50!white, mark = none, line width = 0.5pt] table {data/rate region/S14/S14_Rate_Region_Two_Users_LDM_SDR_Hybrid_2Ltx.txt};
    
    \node[text width = 0.2cm, align = center, anchor = north] at (8.65, 9) {\footnotesize (d)};
    
    \node[draw, rectangle, minimum size = 0.5cm, gray] at (0.75,1.25) {$ \scriptscriptstyle \theta = \frac{4 \pi}{9} $};

     \coordinate (spypoint6) at (axis cs:7.95,5.2);
     \coordinate (magnifyglass6) at (axis cs:5.5,4);
     \spy [gray, size=1.1cm, magnification=8] on (spypoint6) in node[fill=white] at (magnifyglass6); 
     \node [below = 14pt of magnifyglass6,align=center] {\tiny \texttt{RS|FD}};
     \node [below = 18pt of magnifyglass6,align=center] {\tiny \texttt{LDM|FD}};
     \node [below = 22pt of magnifyglass6,align=center] {\tiny \texttt{RS|H|4}};
     \node [below = 26pt of magnifyglass6,align=center] {\tiny \texttt{LDM|H|4}};
        
     \coordinate (spypoint7) at (axis cs:2.1,7.3);
   	 \coordinate (magnifyglass7) at (axis cs:2.1,5);
     \spy [gray, size=1.1cm, magnification=8] on (spypoint7) in node[fill=white] at (magnifyglass7);    
     \node [below = 14pt of magnifyglass7,align=center] {\tiny \texttt{RS|H|2}};
     \node [below = 18pt of magnifyglass7,align=center] {\tiny \texttt{LDM|H|2}};

  \end{groupplot}
\end{tikzpicture}
\end{minipage}
\caption{Two-user rate region for fully-digital (\texttt{FD}) and hybrid (\texttt{H}) precoders.}
\label{fig_rate_region}
\vspace{-2mm}
\end{figure*}
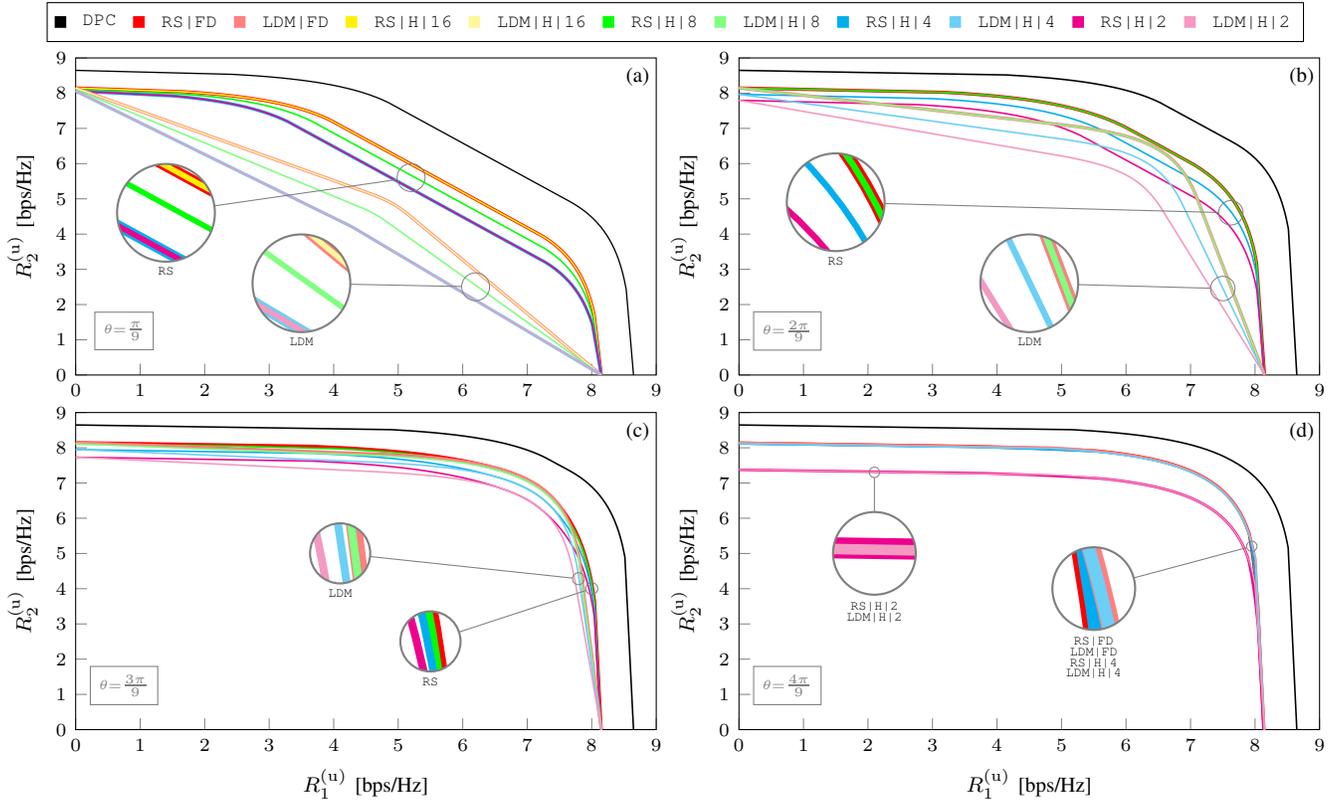

\subsection{Convergence} \label{simulations_convergence}
We compare the convergence of \texttt{FALCON} and \texttt{WMMSE} for a random realization of \texttt{Case C} in Table \ref{t1}, which is the most favorable scenario for \texttt{WMMSE}. Fig. \ref{fig_convergence}a shows the evolution of the unicast rates when $ P^{(0)}_\mathrm{m} = 0.90 P_\mathrm{tx} $. We observe that \texttt{FALCON} converges $ 2 - 3 $ times faster than \texttt{WMMSE} for hybrid and fully-digital precoders. Since the performance of \texttt{WMMSE} relies on a initial point, in Fig. \ref{fig_convergence}b we evaluate more cases for the fully-digital precoder using MRT (in \cite{joudeh2016:rate-splitting-sum-rate-maximization-partial-csit} it is shown that MRT with SVD leads to faster convergence in high SNR regime as in this case). In Fig. \ref{fig_convergence}b, although $ P^{(0)}_\mathrm{m} = 0.80 P_\mathrm{tx} $ produces faster convergence among other choices, it does not surpass \texttt{FALCON}, which only requires $ 8 $ iterations to attain convergence. Further, from Table \ref{t1}, using $ P^{(0)}_\mathrm{m} = 0.80 P_\mathrm{tx} $ only produces $ 69 \% $ of feasible solutions. Also, note that while $ P^{(0)}_\mathrm{m} = 0.99 P_\mathrm{tx} $ produces a feasible solution in $ 94 \%$ of the cases, this choice makes \texttt{WMMSE} converge to a less optimal point. Finding performant initial feasible points for \texttt{WMMSE} is impractical as either infeasibility certificates are returned or the convergence is impacted. In particular, we observe that \texttt{WMMSE} experiences the following trade-off: \textit{large initial power allocation for $ P^{(0)}_\mathrm{m} $ leads to higher likelihood of producing feasible solutions but causes extremely slow convergence (optimality may even be affected) while small initial power allocations may produce faster convergence but increases the likelihood of resulting in infeasible solutions.}

\subsection{Two-user rate region} \label{simulations_two_user_region}
We evaluate the performance of RS-NOUM and LDM-NOUM using \texttt{FALCON} with fully-digital (\texttt{FD}) and projection-based hybrid (\texttt{H}) precoders considering the same settings as in \cite{mao2018:rate-splitting-noma-unicast-multicast}, namely $ K = 2 $, $ N_\mathrm{tx} = 4 $, $ C^\mathrm{th}_0 = 0.5 $ bps/Hz and channels $ \mathbf{h}_1 = \left[ 1, 1, 1, 1 \right]^H $, $ \mathbf{h}_2 = \left[ 1, e^{j \theta}, e^{j 2 \theta}, e^{j 3 \theta} \right]^H $. For the hybrid precoder, we assume that $ N^\mathrm{RF}_\mathrm{tx} = K = 2 $ with four degrees of quantization, $ L_\mathrm{tx} = \left\lbrace 2, 4, 8, 16 \right\rbrace $. To solve the LDM case via \texttt{FALCON}, we enforce $ C_k = 0 $, $ \forall k \in \mathcal{K} $. Fig. \ref{fig_rate_region}a shows the case when the channels are highly correlated ($ \theta = \pi / 9 $). We observe that RS outperforms LDM due to its capability to manage interference, in particular in this challenging scenario. We also note that for RS and LDM, the hybrid precoder with $ L_\mathrm{tx} = 16 $ (i.e., 4 bits) has the same performance as a fully-digital precoder. The performance when $ L_\mathrm{tx} = 2 $ and $ L_\mathrm{tx} = 4 $ is the same due to quantization that has produced the same analog precoder. Through Fig. \ref{fig_rate_region}b, Fig. \ref{fig_rate_region}c, Fig. \ref{fig_rate_region}d, the channel correlation among users is reduced by increasing $ \theta $. As expected, LDM approaches the performance of RS as interference becomes less detrimental. Interestingly, the phase resolution of the hybrid precoder becomes less relevant as the correlation between channels decreases. For instance, in Fig. \ref{fig_rate_region}d with $ L_\mathrm{tx} = 4 $ (i.e., 2 bits) the hybrid and fully-digital precoders attain the same performance.

\subsection{Comparing hybrid precoder designs} \label{simulations_hybrid_precoder_designs}
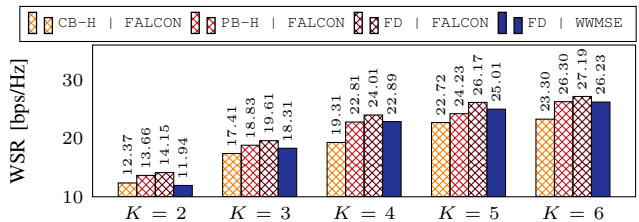
\begin{figure}[!t]
\centering	
\begin{tikzpicture}
	\begin{axis}
    [
		ybar = 0,
		ymin = 10,
		ymax = 36,
		width = 8.8cm,
		height = 3.6cm,
		bar width = 7pt,
		tick align = inside,
		ytick style = {draw = none},
		xtick style = {draw = none},
		x label style={align=center, font=\footnotesize,},
		ylabel = {\footnotesize WSR [bps/Hz]},
		y label style={at={(-0.1,0.5)}, font=\footnotesize,},
		symbolic x coords = {K2, K3, K4, K5, K6},
		xticklabels = {$ K = 2 $, $ K = 3 $, $ K = 4 $, $ K = 5 $, $ K = 6 $},
		x tick label style = {text width = 1.2cm, align = center, font = \fontsize{7}{6}\selectfont},
		y tick label style = {text width = 0.3cm, align = right, font = \fontsize{7}{6}\selectfont},
		xtick = data,
		enlarge y limits = {value = 0.0, upper},
		enlarge x limits = 0.15,
		nodes near coords,
		nodes near coords align = {vertical},
		every node near coord/.append style={rotate=90, anchor=east, font=\tiny, color = white, xshift = 1, /pgf/number format/precision=2, /pgf/number format/fixed zerofill},
		legend columns = -1,
		legend pos = north east,
		legend style={at={(-0.135,1.02)},anchor=south west, font=\fontsize{6}{5}\selectfont, text height=0.02cm, text depth = .ex, fill = none, /tikz/every even column/.append style={column sep=0.046cm}},
		]
 	
		\addlegendimage{black, fill = orange, mark=none, line width=0.5pt, pattern=crosshatch, pattern color = orange},		
		\addlegendentry{\texttt{CB-H | FALCON}},
	
		\addlegendimage{black, fill = dcolor5, mark=none, line width=0.5pt, pattern=crosshatch, pattern color = dcolor5},
		\addlegendentry{\texttt{PB-H | FALCON}},
	
		\addlegendimage{black, fill = dcolor6, mark=none, line width=0.5pt, pattern=crosshatch, pattern color = dcolor6},
		\addlegendentry{\texttt{FD | FALCON}},
			
		\addlegendimage{black, fill = dcolor1, mark=none, line width=0.5pt},
		\addlegendentry{\texttt{FD | WWMSE}},


		\addplot[fill = orange, every node near coord/.append style={color = black, xshift = -2, rotate=0, anchor=west,}, pattern=crosshatch, pattern color = orange] coordinates {(K2, 12.3699) (K3, 17.4081) (K4, 19.3086) (K5, 22.7207) (K6, 23.2996)}; 
		
		\addplot[fill = dcolor5, every node near coord/.append style={color = black, xshift = -2, rotate=0, anchor=west,}, pattern=crosshatch, pattern color = dcolor5] coordinates {(K2, 13.6626) (K3, 18.8315) (K4, 22.8102) (K5, 24.2327) (K6, 26.3001)}; 
		
		\addplot[fill = dcolor6, every node near coord/.append style={color = black, xshift = -2, rotate=0, anchor=west,}, pattern=crosshatch, pattern color = dcolor6] coordinates { (K2, 14.1459) (K3, 19.6111) (K4, 24.0143) (K5, 26.1693) (K6, 27.1904)}; 
		
		\addplot[fill = dcolor1, every node near coord/.append style={color = black, xshift = -2, rotate=0, anchor=west,},] coordinates { (K2, 11.9383) (K3, 18.3145) (K4, 22.8928) (K5, 25.0132) (K6, 26.2294)}; 
		
	\end{axis}
\end{tikzpicture}
\caption{Performance comparison of hybrid precoders}
\label{fig_comparison_hybrid_precoders}
\end{figure}

We compare the performance of \texttt{FALCON} in RS-NOUM with the two hybrid precoder designs described in Section \ref{design_hybrid_precoder}. In this scenario, we assume that $ N_\mathrm{tx} = 8 $, $ C^\mathrm{th}_0 = 1.5 $ (as in \texttt{Case C} of Table \ref{t1} and Table \ref{t2}) for various number of users $ K = \left\lbrace 2, 3, 4, 5, 6 \right\rbrace $. For the PB hybrid (\texttt{PB-H}) precoder we assume that $ L_\mathrm{tx} = 16 $. For CB hybrid (\texttt{CB-H}) precoder, we form the codebook $ \mathcal{V} $ with 128 codewords, where $ N_\mathrm{tx} $ codewords are mutually orthogonal obtained from the discrete Fourier matrix of size $ N_\mathrm{tx} $ and $ 120 $ codewords pseudo-randomly generated. Fig. \ref{fig_comparison_hybrid_precoders} shows that \texttt{PB-H} outperforms \texttt{CB-H} through all values of $ K $. In addition, we include \texttt{FALCON} and \texttt{WMMSE} using fully-digital precoders. For \texttt{WMMSE}, we have assumed $ P^{(0)}_\mathrm{m} = 0.80 P_\mathrm{tx} $ because in Section \ref{simulations_convergence} we showed that this initial value does not affect performance substantially while allowing high convergence speed (although the feasibility ratio is conditioned).

\subsection{Computational complexity} \label{simulations_computational_complexity}
The computational complexity per iteration of \texttt{WMMSE} and \texttt{FALCON} with fully-digital (\texttt{FD}) or hybrid (\texttt{H}) precoders is shown in Table \ref{table_computational_complexity}. The complexity of designing the analog component $ \mathbf{F} $ has been considered negligible due to simple inner products involved. Since the complexity per iteration is not representative of the convergence behavior of a scheme, we show in Table \ref{table_convergence_time} the time required for convergence and the feasibility ratio. We evaluate \texttt{FALCON} and \texttt{WMMSE} using fully-digital precoders for the same scenario described in Section \ref{simulations_hybrid_precoder_designs}. We observe that while \texttt{FALCON} has higher computational complexity per iteration (see Table \ref{table_computational_complexity}), it converges up to $ 100\% $ times faster (in terms of the execution time) and in addition, it returns more feasible solutions than \texttt{WMMSE}.

\begin{table}[t!]
	\vspace{1mm}	 
	\scriptsize
	\caption{Computational complexity per iteration}
	\centering
	\begin{tabular}{|l|l|l|}
		\hline
		\centering
		{\bf \centering{~~~~~~~~~~Description}} & {\bf \centering{~Notation}} & {\bf \centering ~~~~~Complexity }	\\ 
		\hline
		\hline
		{\scriptsize Weights of \texttt{WWMSE} with \texttt{FD} design} & $ \mathcal{C}^{(\mathrm{w})}_\texttt{FD-WMMSE} $ & $ \mathcal{O} \left(  K^2 N_\mathrm{tx} \right) $  \\ 
		\hline
		{\scriptsize Precoders of \texttt{WWMSE} with \texttt{FD} design} & $ \mathcal{C}^{(\mathrm{p})}_\texttt{FD-WMMSE} $  & $ \mathcal{O} \left(  K^{3.5} \left[ N_\mathrm{tx} \right]^{3.5} \right) $ \\ 
		\hline
		{\scriptsize Weights of \texttt{WWMSE} with \texttt{H} design} & $ \mathcal{C}^{(\mathrm{w})}_\texttt{H-WMMSE} $ & $ \mathcal{O} \left(  K^2 N^\mathrm{RF}_\mathrm{tx} \right) $  \\ 
		\hline
		{\scriptsize Precoders of \texttt{WWMSE} with \texttt{H} design} & $ \mathcal{C}^{(\mathrm{p})}_\texttt{H-WMMSE} $  & $ \mathcal{O} \left(  K^{3.5} \left[ N^\mathrm{RF}_\mathrm{tx} \right]^{3.5} \right) $ \\ 
		\hline
		{\scriptsize \texttt{FALCON} with \texttt{FD} design} & $ \mathcal{C}_\texttt{FD-FALCON} $ & $ \mathcal{O} \left( K^{3} \left[ N_\mathrm{tx} \right]^{6} \right) $  \\ 
		\hline
		{\scriptsize \texttt{FALCON} with \texttt{H} design} & $ \mathcal{C}_\texttt{H-FALCON} $ & $ \mathcal{O} \left( K^{3} \left[ N^\mathrm{RF}_\mathrm{tx} \right]^{6} \right) $  \\ 
		\hline
	\end{tabular}
	\label{table_computational_complexity}
\end{table}
\begin{table}[t!]
	\setlength{\tabcolsep}{0.42em} 
	\scriptsize
	\caption{Convergence time}
	\centering
	\begin{tabular}{|c|c|c|c|c|c|}
		\hline 
		\centering
		\multirow{2}{*}{Scheme} & \multicolumn{5}{c|}{Number of users} \\ 
		\cline{2-6} 
								& {$ K = 6 $} & {$ K = 5 $}  & {$ K = 4 $} & {$ K = 3 $} & {$ K = 2 $} \\ 
		\hline
		\hline
		\texttt{FALCON} & $ 93.6 \text{s} | 91\%  $ & $ 40.1 \text{s} | 93\% $ & $ 33.8 \text{s} | 96\% $ & $ 31.7 \text{s} | 100\% $ & $ 37.5 \text{s} | 100\% $\\ 
		\hline
		\texttt{WWMSE} & $ 116.9 \text{s} | 67\% $  & $ 75.9 \text{s} | 82\% $ & $ 66.3 \text{s} | 75\% $ & $ 45.1 \text{s} | 79\% $ & $ 31.4 \text{s} | 62\% $ \\ 
		\hline
	\end{tabular}
	\label{table_convergence_time}
\end{table}

\section{Conclusion} \label{conclusion}
We have investigated non-orthogonal unicast multicast transmissions by means of rate-splitting with fully-digital and hybrid precoders. We considered the weighted sum-rate maximization problem with a minimum multicast QoS requirement. We proposed \texttt{FALCON} based on sequential parametric optimization, which is shown to outperform \texttt{WMMSE} under a variety of scenarios. We showed that \texttt{FALCON} converges to local optimum of the nonconvex problem. Although \texttt{FALCON} has higher complexity per iteration than \texttt{WMMSE}, it converges faster within a few iterations and does not require an initial feasible point, which impacts performance. Further, \texttt{FALCON} is a viable option for designing hybrid precoders with rate-splitting since the complexity scales with the number of RF chains, which in general, is relatively small. In addition, for the two-user case, we noticed that a quantization scheme with 4 bits is sufficient for guaranteeing performance equal to that of a fully-digital precoder. Moreover, as the user channels become less correlated, the quantization granularity of the analog precoder is less relevant.

\section*{Acknowledgment} \label{acknowledgment}
This research is funded by the Deutsche Forschungsgemeinschaft (DFG) within the B5G-Cell project in SFB 1053 MAKI, and the LOEWE initiative (Hesse, Germany) within the emergenCITY centre.

\setcounter{equation}{0}
\renewcommand{\theequation}{A.\arabic{equation}}

\bibliographystyle{IEEEtran}
\bibliography{ref}

\end{document}